\documentclass{emulateapj}
\usepackage{apjfonts}
\usepackage{epsfig}
\usepackage{lineno}

\journalinfo{The Astrophysical Journal}
\slugcomment{Accepted for publication, 2011 Nov 17}

\shortauthors{CAMILO ET AL.}
\shorttitle{PSR~J2030+3641: MIDDLE-AGED RADIO AND GAMMA-RAY PULSAR}

\begin{document}

\def\fermi{{\em Fermi}}
\def\ergs{\,erg\,s$^{-1}$}
\def\pccc{\,pc\,cm$^{-3}$}

\def\fgl{1FGL~J2030.0+3641}
\def\psr{PSR~J2030+3641}

\title{PSR~J2030+3641: radio discovery and gamma-ray study of
a middle-aged pulsar in the now identified Fermi-LAT source
1FGL~J2030.0+3641}

\author{
F.~Camilo\altaffilmark{1,2},
M.~Kerr\altaffilmark{3,4,5},
P.~S.~Ray\altaffilmark{6,7},
S.~M.~Ransom\altaffilmark{8},
S.~Johnston\altaffilmark{9},
R.~W.~Romani\altaffilmark{3},
D.~Parent\altaffilmark{10},
M.~E.~DeCesar\altaffilmark{11,12},
A.~K.~Harding\altaffilmark{11},
D.~Donato\altaffilmark{12,13},
P.~M.~Saz~Parkinson\altaffilmark{14},
E.~C.~Ferrara\altaffilmark{11},
P.~C.~C.~Freire\altaffilmark{15},
L.~Guillemot\altaffilmark{15},
M.~Keith\altaffilmark{9},
M.~Kramer\altaffilmark{15,16},
and K.~S.~Wood\altaffilmark{6}
}

\altaffiltext{1}{Columbia Astrophysics Laboratory, Columbia University, New York, NY 10027, USA}
\altaffiltext{2}{email: fernando@astro.columbia.edu}
\altaffiltext{3}{W. W. Hansen Experimental Physics Laboratory, Kavli Institute for Particle Astrophysics and Cosmology, Department of Physics and SLAC National Accelerator Laboratory, Stanford University, Stanford, CA 94305, USA}
\altaffiltext{4}{email: kerrm@stanford.edu}
\altaffiltext{5}{Einstein Fellow}
\altaffiltext{6}{Space Science Division, Naval Research Laboratory, Washington, DC 20375-5352, USA}
\altaffiltext{7}{email: Paul.Ray@nrl.navy.mil}
\altaffiltext{8}{National Radio Astronomy Observatory (NRAO), Charlottesville, VA 22903, USA}
\altaffiltext{9}{CSIRO Astronomy and Space Science, Australia Telescope National Facility, Epping NSW 1710, Australia}
\altaffiltext{10}{Center for Earth Observing and Space Research, College of Science, George Mason University, Fairfax, VA 22030, resident at Naval Research Laboratory, Washington, DC 20375, USA}
\altaffiltext{11}{NASA Goddard Space Flight Center, Greenbelt, MD 20771, USA}
\altaffiltext{12}{Department of Physics and Department of Astronomy, University of Maryland, College Park, MD 20742, USA}
\altaffiltext{13}{Center for Research and Exploration in Space Science and Technology (CRESST) and NASA Goddard Space Flight Center, Greenbelt, MD 20771, USA}
\altaffiltext{14}{Santa Cruz Institute for Particle Physics, Department of Physics and Department of Astronomy and Astrophysics, University of California at Santa Cruz, Santa Cruz, CA 95064, USA}
\altaffiltext{15}{Max-Planck-Institut f\"ur Radioastronomie, Auf dem H\"ugel 69, 53121 Bonn, Germany}
\altaffiltext{16}{Jodrell Bank Centre for Astrophysics, School of Physics and Astronomy, The University of Manchester, M13 9PL, UK}

\begin{abstract}
In a radio search with the Green Bank Telescope of three unidentified
low Galactic latitude \fermi-LAT sources, we have discovered the
middle-aged pulsar J2030+3641, associated with \fgl\ (2FGL~J2030.0+3640).
Following the detection of gamma-ray pulsations using a radio ephemeris,
we have obtained a phase-coherent timing solution based on gamma-ray and
radio pulse arrival times that spans the entire \fermi\ mission.  With a
rotation period of 0.2\,s, spin-down luminosity of $3\times10^{34}$\ergs,
and characteristic age of 0.5\,Myr, \psr\ is a middle-aged neutron star
with spin parameters similar to those of the exceedingly gamma-ray-bright
and radio-undetected Geminga.  Its gamma-ray flux is 1\% that of Geminga,
primarily because of its much larger distance, as suggested by the
large integrated column density of free electrons, $\mbox{DM}=246$\pccc.
We fit the gamma-ray light curve, along with limited radio polarimetric
constraints, to four geometrical models of magnetospheric emission, and
while none of the fits have high significance some are encouraging and
suggest that further refinements of these models may be worthwhile.
We argue that not many more non-millisecond radio pulsars may be
detected along the Galactic plane that are responsible for LAT sources,
but that modified methods to search for gamma-ray pulsations should be
productive --- \psr\ would have been found blindly in gamma rays if only
$\ga 0.8$\,GeV photons had been considered, owing to its relatively flat
spectrum and location in a region of high soft background.

\end{abstract}


\section{Introduction} \label{sec:intro} 

Three years after the launch of the \textit{Fermi Gamma-ray
Space Telescope}, pulsations have been detected with its Large
Area Telescope (LAT) from at least 100 rotation-powered neutron
stars \citep[e.g.,][]{abdo11}, increasing by more than an order of
magnitude the number of pulsars known to emit above 0.1\,GeV\footnote{See
https://confluence.slac.stanford.edu/display/GLAMCOG/Public+List+of+LAT-Detected+Gamma-Ray+Pulsars
for an up-to-date list.}.  Nearly 20 of these are millisecond
pulsars (MSPs) known independent from \fermi, for which rotational
ephemerides obtained from radio observations were used to fold the
sparse gamma-ray photons \citep[e.g.,][]{abdo10,abdo12}.  Another 34
have been discovered in direct pulsation searches of the gamma-ray data
\citep[e.g.,][]{abdo2,pga+11,sdz+10}.

Among the known non-MSP gamma-ray pulsar sample, about 75\% have large
values of spin-down luminosity ($\dot E > 10^{35}$\ergs); the remaining
17 have lower $\dot E$ and are older, with characteristic age $\tau_c =
P/(2\dot P) \sim 0.1$--1\,Myr.  Five of these are also radio emitters
\citep[two are exceedingly faint and were detected only after the
gamma-ray discoveries;][]{crr+09,pga+11}.  Almost all of these middle-aged
pulsars are also known or expected to be nearby, at $d \la 1$\,kpc,
although obviously many more similar pulsars exist at larger distances.
In principle several of them could be detected in gamma rays despite
a relatively low $\dot E$, because the efficiency for conversion of
rotational kinetic energy into gamma rays, $\eta \equiv L_{\gamma}/\dot
E$, apparently increases with decreasing $\dot E$ and may approach 100\%
for $\dot E \approx 10^{33-34}$\ergs \citep[see][]{abdo11,aro96b,mh03}.

In order to improve our understanding of pulsar emission mechanisms and
evolution, it is important to identify more gamma-ray-emitting neutron
stars, particularly in under-represented groups such as middle-aged
pulsars.  The Fermi LAT First Source Catalog (1FGL), based on 11 months
of survey data and containing more than 600 unidentified sources
\citep{abdo14}, provides a path toward such discoveries \citep[as
does now 2FGL;][]{abdo17}.  Many unidentified LAT sources are being
surveyed within the context of a ``pulsar search consortium'' that
aims to efficiently organize this collaborative radio and gamma-ray
work \citep[see, e.g.,][]{rp11}.  Spectacularly, many radio MSPs have
recently been discovered in such searches at high Galactic latitudes
\citep[$|b|>5\arcdeg$; e.g.,][]{cgj+11,kjr+11,rrc+11}.  Here we report
on the first pulsar to be discovered at low Galactic latitude that is
responsible for a formerly unidentified 1FGL source.  \psr\ in \fgl\
(also 2FGL~J2030.0+3640), discovered with the NRAO Green Bank Telescope
(GBT), is also the first non-MSP gamma-ray pulsar identified in radio
searches of 1FGL sources.

\section{Observations and Results} \label{sec:obs} 

\subsection{Radio Searches} \label{sec:search}

Consistent with the properties of known gamma-ray pulsars, our first
1FGL search targets are non-variable and have spectra consistent
with exponentially-cutoff power laws.  In this initial small radio
survey at the GBT we aimed to search along the Galactic plane,
indicating a relatively high search frequency ($\ga 1$\,GHz) in order
to minimize deleterious propagation effects in the interstellar medium
and the background temperature due to Galactic synchrotron emission.
At the 2\,GHz frequency used, the GBT beam is $3'$ HWHM, and the 1FGL
sources to be searched should therefore, ideally, have $r_{95}$ (95\%
confidence level error radii) below that, after accounting for the
combined statistical and estimated systematic positional uncertainties.
These are strict criteria, and we searched a total of only three sources.
For each we provide ($\alpha,\delta,\Delta_{\theta},r_{95}$), which
are respectively the R.A. and decl. of the {\em observed} position,
its offset from the 1FGL position (we observed in late 2009, based on an
interim version of the catalog), and $r_{95}$ for the gamma-ray source:
1FGL~J0224.0+6201c ($36\fdg00,62\fdg04,1',4'$), where the ``c'' indicates
that the measured properties of this source (including position) may not
be reliable; 1FGL~J1746.7--3233 ($266\fdg70,-32\fdg61,3',3'$); and \fgl\
($307\fdg52,36\fdg68,1',3'$).

We observed each source for 1\,hr using the GUPPI
spectrometer\footnote{https://wikio.nrao.edu/bin/view/CICADA/GUPPiUsersGuide}
at a central frequency of 2\,GHz.  Each of 512 polarization-summed
frequency channels, spanning a bandwidth of 800\,MHz, were sampled
every 0.164\,ms before writing to disk.  The flux density limit of
these searches, converted to a standard search frequency of 1.4\,GHz, was
approximately 0.03\,mJy, for an assumed pulsar duty cycle of 10\% and spin
period larger than a few tens of milliseconds.  We analyzed the data with
standard pulsar search techniques implemented in PRESTO \citep{ran01},
selecting trial dispersion measures so as to maintain ideal sensitivity
up to twice the maximum Galactic DM predicted in each direction by
the \citet{cl02} electron distribution model.  The data sets retained
significant sensitivity to (possibly binary) MSPs up to $\mbox{DM}
\approx 100$\pccc.  We therefore did the analysis in two passes,
including one where we allowed for a modest amount of unknown constant
acceleration (parameterized in PRESTO by $\mbox{zmax}=50$).  In the \fgl\
data, collected on 2009 November 27, we discovered on Christmas day an
obvious pulsar with period $P=0.200$\,s and $\mbox{DM}=247$\pccc.

\subsection{Radio Timing and Polarimetry of \psr} \label{sec:timing}

We began regular timing observations of the new pulsar at GBT on
2010 January 8.  Observing parameters were identical to those used in
the search observation, although with much shorter integration times,
typically 5 minutes.  In this manner we measured pulse times of arrival
(TOAs) on 28 separate days through 2011 April 11.  Using these with
TEMPO\footnote{http://www.atnf.csiro.au/research/pulsar/tempo} we
obtain a phase-connected rotational ephemeris with 0.3\,ms rms residual.
After 5 months of radio timing, the solution was sufficient to yield an
initial detection of gamma-ray pulsations (see Section~\ref{sec:gamma}).
The DM measurement was improved with the aid of one additional observation
centered at 1.5\,GHz, and implies an uncertainty in the alignment of
radio and gamma-ray pulses of only 0.7\,ms.

\begin{figure}[t]
\centerline{
\hfill
\includegraphics[scale=0.39]{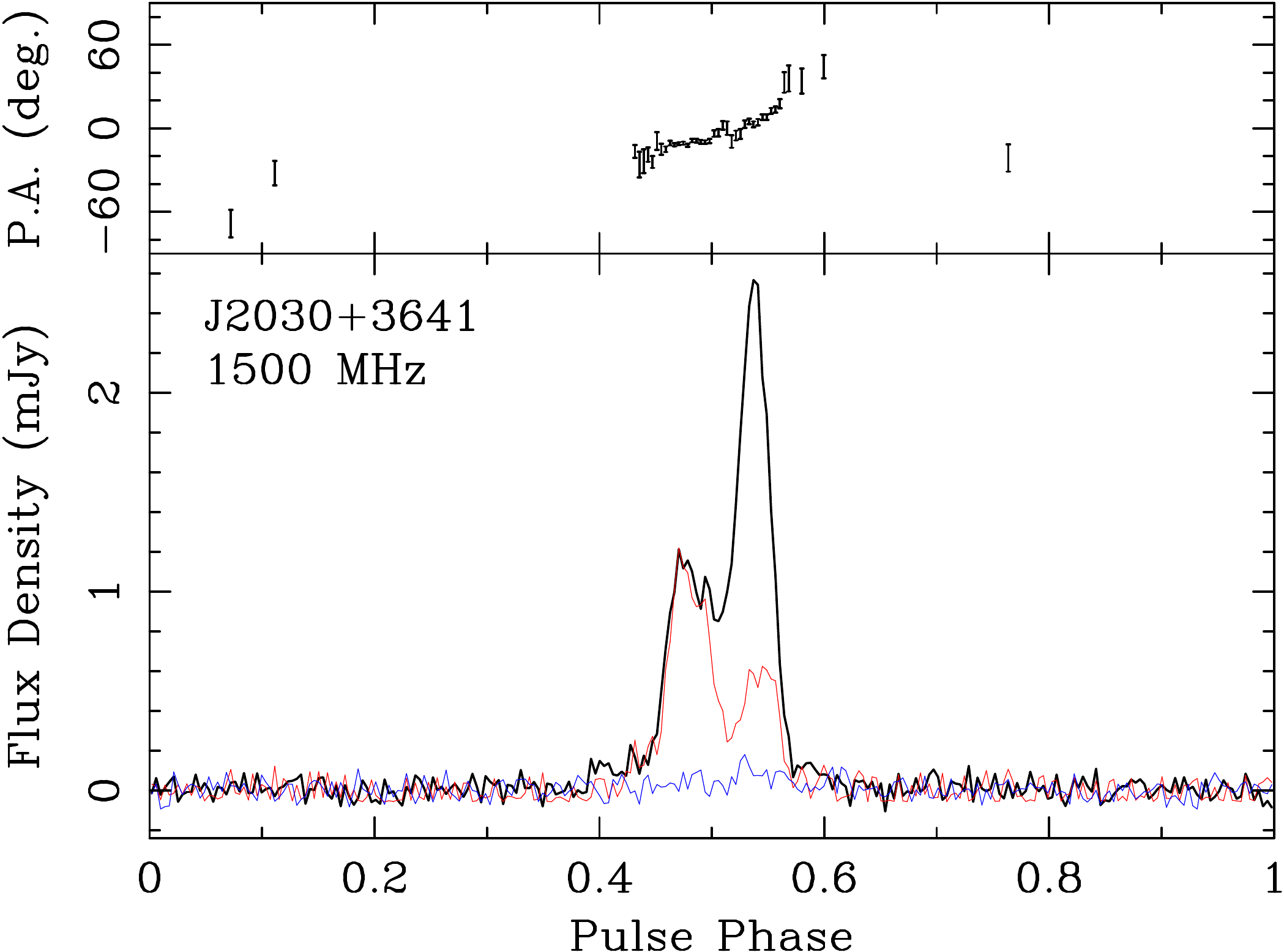}
\hfill
}
\caption{\label{fig:pol}
Polarimetric observation of \psr\ at 1.5\,GHz with the GBT.  In the
lower panel the total-intensity profile is represented by the black
trace, red corresponds to linear polarization, and blue to circular.
The profiles, based on a 4.5\,hr integration, are displayed with a
resolution of 256 bins and arbitrary phase.  In the upper panel the
position angle of linear polarization is displayed for bins in which the
linear signal-to-noise ratio exceeds 2.2, and the PAs have been rotated
to the pulsar frame using the measured $\mbox{RM}=+514$\,rad\,m$^{-2}$.
The full pulse phase corresponds to $P=0.200$\,s.  }
\end{figure}

In order to determine the polarization characteristics of \psr, we have
used GUPPI to make three flux-calibrated full-Stokes observations of
the pulsar, one each at central frequencies of 0.82\,GHz (lasting for
2.3\,hr), 1.5\,GHz (4.5\,hr), and 2\,GHz (1.0\,hr).  We analyzed
the data with PSRCHIVE \citep{hvm04}, and show the resulting
1.5\,GHz profiles in Figure~\ref{fig:pol}.  The profiles at the
other two frequencies are comparable, with a possible slight hint of
interstellar scattering visible at 0.8\,GHz.  The Faraday rotation
measure is $\mbox{RM}=+514$\,rad\,m$^{-2}$, which implies an average
electron-weighted Galactic magnetic field strength along the line of
sight of $2.6\,\mu$G.  The radio spectral index, based on the three
flux density measurements, is $\alpha = -1.7\pm0.1$ (where $S_{\nu}
\propto \nu^{\alpha}$; see Table~\ref{tab:parms}).

\subsection{Gamma-ray Study of \psr} \label{sec:gamma}

We used ``Pass 6 diffuse''-class \fermi\ LAT events (having the
highest probability of being gamma-ray photons), excluding those
with zenith angles $>100\arcdeg$ to minimize gamma-ray emission from
the Earth's limb.  The events were analyzed using the LAT \textit{Science
Tools}\footnote{http://fermi.gsfc.nasa.gov/ssc/data/analysis/scitools/overview.html},
and photon phases were calculated using TEMPO2 \citep{hem06} with the
\texttt{fermi} plugin \citep{rkp+11}.

\begin{figure}[t]
\begin{center}
\includegraphics[scale=0.42]{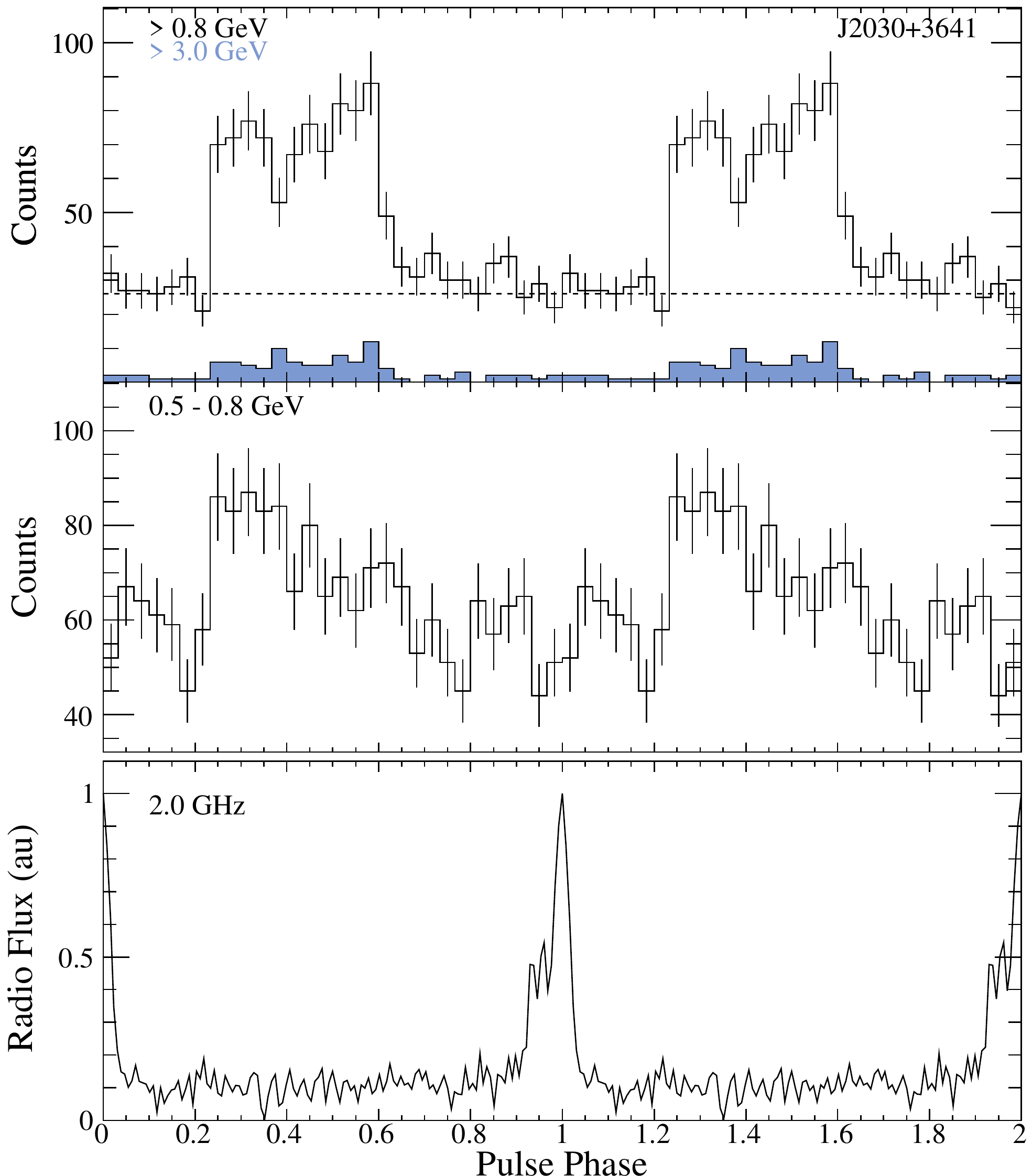}
\end{center}
\caption{\label{fig:prof}
Phase-aligned radio and gamma-ray profiles of \psr.  Bottom panel:
discovery observation at the GBT.  Middle panel: \fermi\ LAT 0.5--0.8\,GeV
profile (significant LAT pulsations are not detected below this energy
band).  Top panel: $>0.8$\,GeV and $>3$\,GeV (lower) LAT profiles.
Each pulse profile is repeated in phase, and displayed with 30 bins per
period in gamma rays and 128 bins in radio.  The dashed line gives the
level of unpulsed emission in the aperture as estimated from the light
curve fits.  }
\end{figure}

We first folded photons detected from \fgl\ based on the initial
radio timing solution for \psr\ (Section~\ref{sec:timing}).  Selecting
photons using guesses for the minimum energy and radius of interest for
extraction, we made a clear detection of pulsations.  Optimal cuts in this
region of very high background were subsequently found with $E>0.8$\,GeV
and within a $0\fdg5$ radius.  We then obtained 20 gamma-ray TOAs from
a total of 1402 photons detected between 2008 August 4 and 2011 April 20
(49 days of integration per TOA), and used TEMPO2 to determine a timing
solution with an rms of 4.0\,ms.  Despite the much higher precision
of the individual radio TOAs, the best overall ephemeris is obtained
by combining the gamma-ray and radio TOAs, owing to the longer span of
the former.  The parameters of this timing solution, in which a small
amount of ``timing noise'' is parameterized by $\ddot \nu$ (where $\nu =
1/P$), are presented in Table~\ref{tab:parms}.

Figure~\ref{fig:prof} shows the gamma-ray pulse profiles in three energy
bands, along with the radio discovery profile.  In order to optimize
each pulsed signal-to-noise ratio, we selected photons within $0\fdg8$,
$0\fdg5$, and $0\fdg3$ of the timing position, respectively for the
0.5--0.8\,GeV, $>0.8$\,GeV, and $>3$\,GeV bands.  For the middle energy
band, the bin-independent H-test \citep{db10}, with a value of 270,
indicates enormous significance.  For each of the other two bands, the
H-test values of approximately 40 correspond to pulsation significances
over $5\,\sigma$.  The light curve at $>0.8$\,GeV consists of two
(conceivably three) overlapping components as observed, e.g., for
PSRs~J1709--4429, J1057--5226, and the CTA~1 pulsar \citep{abdo11},
which have peak separation $\Delta \approx 0.25$.  While the statistics
are poor, the $>3$\,GeV profile is consistent with such a description.
At 0.5--0.8\,GeV, only one peak is visible, which corresponds in phase
with the first peak visible at higher energies.  No pulsed signal was
detected below 0.5\,GeV, despite searches using different apertures.
We fit the unbinned gamma-ray data above 0.8\,GeV with two two-sided
Gaussians, to account for the steep leading and trailing edges of
the pulsed emission. The first peak is offset in phase from the main
radio peak by $\delta = 0.26 \pm 0.02$ with $\mbox{FWHM} = 0.11 \pm
0.04$, while the second peak is offset from the first by $\Delta =
0.30^{+0.03}_{-0.08}$, with $\mbox{FWHM} = 0.18 \pm 0.05$. We have
included systematic error from the modest uncertainty in DM and our
choice of model.

\subsubsection{Spectrum} \label{sec:spect}

To characterize the spectrum of \psr, we analyzed Pass 6 diffuse events
collected between 2008 August 4 and 2011 April 20 selected to have
reconstructed energies between 178\,MeV and 100\,GeV and reconstructed
positions within $10^{\circ}$ of the timing position of \psr.
We excluded the 100--178\,MeV energy bin because the rapidly increasing
effective area in this bin exacerbates the effects of energy dispersion
\cite[][]{aaa+09}.  We removed events with a reconstructed zenith angle
$>100^{\circ}$ and also filtered data in which the aperture approached
the limb of the Earth; these cuts remove most of the contribution of
the bright gamma-ray emission produced by cosmic rays interacting in
the Earth's atmosphere.  To further enhance the signal-to-noise ratio
of \psr, we selected events with rotation phase $0.22<\phi<0.64$ (see
Figure \ref{fig:prof}).  We verified that there was negligible emission
outside of this phase range, so the best-fit values may be taken to
represent the phase-averaged magnetospheric emission.

\psr\ lies in the source-crowded Cygnus region, and extracting its
spectrum requires careful modeling of both neighboring point sources
(including the three bright gamma-ray pulsars J2021+3651, J2021+4026,
and J2032+4127) and the diffuse background.  We included all point
sources in a preliminary version of the 2FGL catalog \citep{abdo17}
within $15\arcdeg$ of the pulsar, as the broad point-spread function
(PSF) at low energy allows sources away from the aperture boundary
to contribute to the observed counts.  Identified pulsars were
modeled with a power law with exponential cutoff, $dN/dE \propto
(E/{\rm GeV})^{-\Gamma}\exp(-E/E_c)$.  In the fits reported below, we
varied all parameters for sources within $7\arcdeg$ of the aperture,
while sources outside of the aperture were held fixed at the catalog
values.  We modeled the Galactic and isotropic diffuse backgrounds using,
respectively, the \textit{gll\_iem\_v02\_P6\_V11\_DIFFUSE} diffuse map and
the \textit{isotropic\_iem\_v02\_P6\_V11\_DIFFUSE} tabulation employed
for the 1FGL spectral analysis; these have been rescaled from those
constructed with the \textit{P6\_V3\_DIFFUSE} instrument response functions
(IRFs).  We allowed the normalizations of the diffuse models to vary in
the fit.

We performed spectral fits with the \textit{pointlike} application
\citep{ker11}.  This software bins the events in both energy and position,
using HEALPix\footnote{http://healpix.jpl.nasa.gov/} to provide position
bins whose energy-dependent extent is always small compared to the PSF,
providing a compact representation of the data with minimal information
loss.  We verified the results with the ScienceTool \textit{gtlike}.
The best-fit values are reported in Table~\ref{tab:parms}, and include
a power-law photon index $\Gamma = 1.1$ and cut-off energy $E_c =
2.0$\,GeV.  These values apply to the profile as a whole, although
Figure~\ref{fig:prof} suggests that the second gamma-ray peak has
a harder spectrum than the first, as is observed for many gamma-ray
pulsars \citep[e.g.,][]{awd+11}.

In addition to the rescaled 1FGL diffuse model, we also
performed the fit with an improved model internal to the
LAT collaboration and with ``Pass 7'' data\footnote{See
http://fermi.gsfc.nasa.gov/ssc/data/analysis/scitools/} using the
\textit{P7SOURCE\_V6} IRFs and the appropriate diffuse model (accompanying
the 2FGL catalog).  These values were consistent with those reported
in Table~\ref{tab:parms}, and the scatter has been used to estimate
systematic errors on the parameters.  We have also included uncertainty
in the instrument's effective area in the systematic errors through the
use of ``bracketing'' IRFs \citep{abdo2009}.

\begin{figure*}[t]
\begin{center}
\includegraphics[scale=0.83]{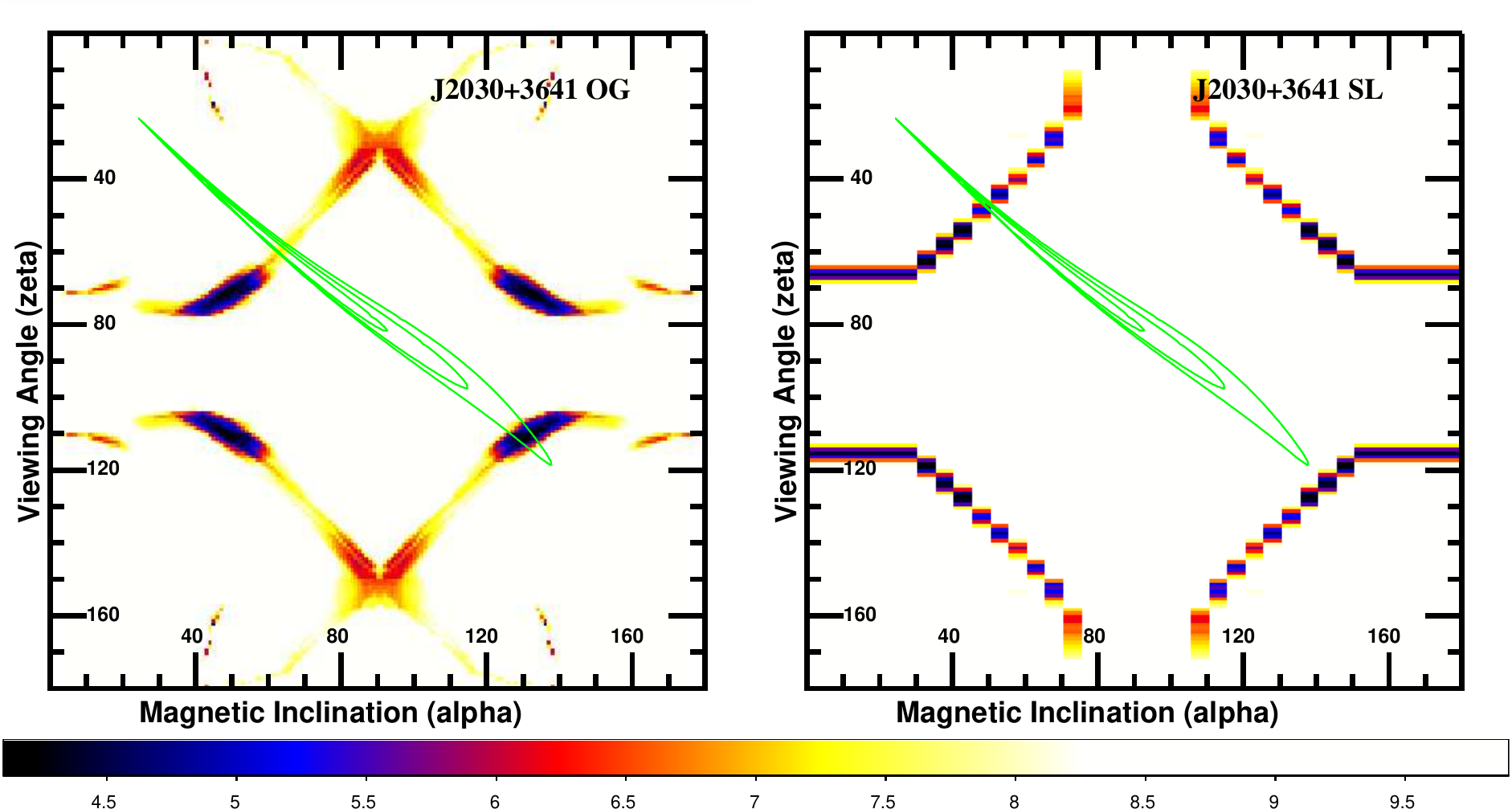}
\end{center}
\caption{\label{fig:chi3}
Surfaces of goodness of fit ($\chi_3$) to the $E>0.8$\,GeV light curve
for the OG (left) and SL (right) outer magnetosphere beaming models.
Dark colors indicate lower (better) values of the fit statistic. The
$\Delta \chi^2 = +1, +2, +3$ contours for the RVM fit to the 1.5\,GHz
polarization data are shown in green.  }
\end{figure*}

\subsection{X-ray Observations} \label{sec:x-ray}

A short \textit{Swift} \citep{gcg+04} observation was obtained
on 2010 May 11 to search for an X-ray counterpart to \psr.
No emission is observed at the pulsar position in this 3.8\,ks X-ray
Telescope \citep{bhn+05} PC mode observation, with a $3\,\sigma$
upper limit of $1.7\times10^{-3}$\,cts\,s$^{-1}$ (0.5--8 keV).
Assuming a power-law spectrum with photon index $\Gamma=1.5$ for
a possible pulsar wind nebula \citep[e.g.,][]{kp08} absorbed by
a column with $N_H = 7.4\times10^{21}$\,cm$^{-2}$ (assuming one
free electron for every 10 neutral hydrogen atoms along the line
of sight), this corresponds to an unabsorbed flux limit of $f_X <
1.3\times10^{-13}$\,erg\,cm$^{-2}$\,s$^{-1}$, or luminosity $L_X <
1.6\times10^{31} (d/1\,{\rm kpc})^2$\ergs.  For a distance of $\approx
2$\,kpc (see Table~\ref{tab:parms}), this corresponds to $L_X/\dot E <
2\times10^{-3}$, which is a reasonably stringent limit, although it would
not be a surprise if the actual X-ray flux of this pulsar were about an
order of magnitude below the current limit \citep[see, e.g.,][]{mdc11}.

\section{Discussion} \label{sec:disc} 

The gamma-ray pulsations from \psr\ unambiguously identify it as the
origin of the gamma rays from \fgl.  We note also that a candidate
TeV source possibly detected by the Milagro Gamma-Ray Observatory
\citep[``C2'' in][]{aab+07} is positionally coincident with \psr.
Its faintness and location near a very bright source precludes
further investigation based only on published results \citep[see
also][]{milagro09}, but the relatively small $\dot E$ and relatively large
age and distance of \psr\ make an association unlikely.  With a DM-derived
distance of 8\,kpc, \psr\ is located toward the Cygnus region, where until
recently very few pulsars were known and where the pulsar distance scale
is highly uncertain.  Recent discoveries suggest that the \citet{cl02}
model greatly overpredicts distances in this direction: PSR~J2032+4127,
located $5\arcdeg$ from \psr, with a DM-derived distance of 3.6\,kpc,
is thought to be located at half that distance \citep{crr+09}; and
PSR~J2021+3651, less than $2\arcdeg$ from \psr, with a model distance of
12\,kpc, is more likely located at 2--4\,kpc \citep{abdo15}.  Also, the RM
for \psr\ is large and positive, very similar to that of PSR~J2021+3651,
and this is suggestive of a location closer than the Perseus arm, at
6\,kpc in this direction \citep[see discussion in][]{hml+06}.  Overall,
we judge that a likely distance for \psr\ is in the range 2--4\,kpc.

\subsection{Geometrical Constraints from Radio Profile} \label{sec:simon}

\citet{jw06} noted that the profiles of pulsars with high $\dot{E}$
share many common features.  In particular, the profiles with two
components generally show (i) two components with equal width, (ii)
the trailing component stronger than the leading component, (iii) close
to 100\% linear polarization, (iv) significant circular polarization
under the trailing component, and (v) flat position angle swings which
appear to steepen at the far trailing edge of the profile.  The profile
of \psr\ conforms in some ways to these generalizations.  The profile is
symmetrical with the trailing edge brighter. The PA swing is remarkably
flat with a hint of steepening at the trailing edge.  However, although
the leading component is highly polarized, the trailing edge is not,
unlike the majority of the high $\dot{E}$ pulsars.  \citet{wj08} pointed
out that the transition between pulsars with low linear polarization and
those which are highly polarized occurs at $\dot{E}\sim 10^{34.5}$\ergs.
\psr\ could be an example of a pulsar in transition from high to low
polarization.

The interpretation of these profiles by \cite{jw06} was that the magnetic
pole crossing occurs at the (symmetry) center of the profile.  The PA
swing is significantly shifted to later phase because of aberration
\citep{bcw91}, leading to emission heights close to 1000\,km.  This may
also be the case in \psr.  If we assume that the inflexion point of the
PA swing is the rise in the PA curve some $20^{\circ}$ after the profile
center in Figure~\ref{fig:pol}, this implies an emission height of $\sim
800$\,km (close to 10\% of the light cylinder), similar to those seen
in other high $\dot{E}$ pulsars.  We also note that a height of $\sim
800$\,km would in turn imply an overall pulse width of some $50^{\circ}$,
similar to the observed value.  This further suggests that the inclination
between the magnetic and rotation axes cannot be too far from orthogonal.

\subsection{Light Curve Fitting and Model Implications} \label{sec:roger}

We can use the gamma-ray light curve and polarimetric measurements to
test the predictions of magnetosphere beaming models. Such tests are most
constraining when the magnetic inclination angle $\alpha$ and the viewing
angle $\zeta$ are well determined.  The polarization data provide some
geometrical constraints.  Since the lower frequency data may be affected
by interstellar scattering, our primary constraints are based on the
1.5\,GHz data.  We use PA values from the native 512 observational
phase bins with $\sigma_{\rm PA}<10^\circ$ and phases $-0.1 <\phi <
0.05$ (relative to the intensity maximum; see Figure~\ref{fig:pol}),
and fit to a standard rotating vector model (RVM) polarization sweep.
The maximum sweep rate is d$\Psi$/d$\phi |$max = 1.4 and one obtains
good fits in a band of the $\alpha$--$\zeta$ plane, with a best fit
$\chi^2 = 37.5$ for 55 degrees of freedom (suggesting the PA errors
are somewhat overestimated).  Figure~\ref{fig:chi3} shows the RVM fit
confidence regions in this plane, after correcting for the RVM sign
convention problem \citep{ew01}.  The phase of maximum sweep is not
well constrained, but the best fits place it well after the radio peak
at $\phi\approx 0.06$.  The fit implies $\alpha < 140^\circ$, with best
fit values $<90^\circ$.  If we simultaneously fit the 1.5\,GHz points
and the $-0.1 <\phi < 0.02$ 0.8\,GHz data, larger $\alpha > 90^\circ$
are preferred, but this preference is controlled largely by the trailing
edge of the sweep where scattering distortions may be present.

\begin{figure}[t]
\begin{center}
\includegraphics[scale=0.51]{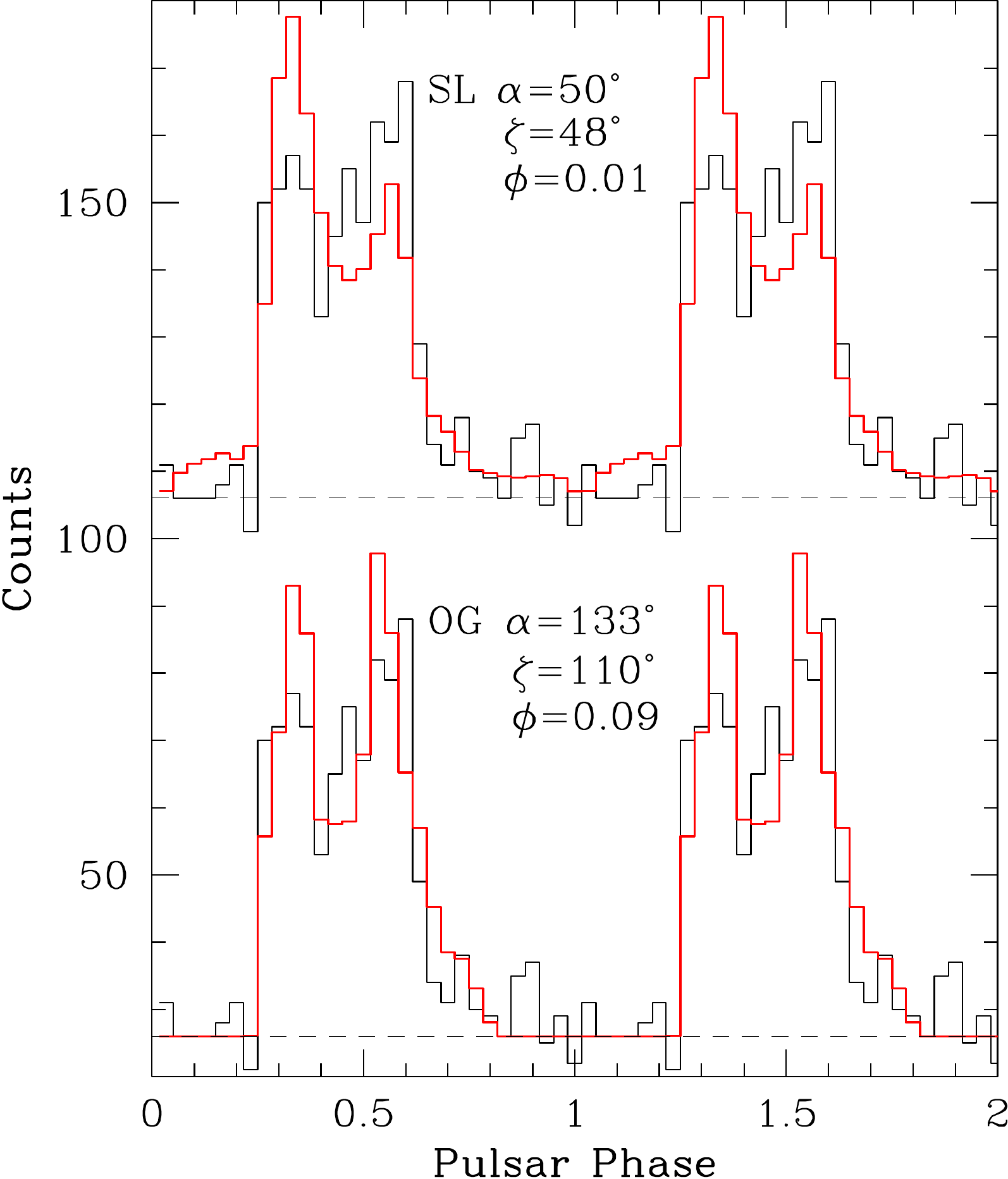}
\end{center}
\caption{\label{fig:lc}
Model light curves (red) at the best-fit $\alpha$ and $\zeta$ consistent
with the RVM PA constraints compared to the $E > 0.8$\,GeV LAT light
curve (black histograms). The offset $\phi$ indicates the phase of
the closest approach of the magnetic dipole axis at the star surface.
The SL curves have been offset by +80 counts/bin.  }
\end{figure}

Following \citet{rw10}, we next compare the gamma-ray light curve to
three geometrical models of outer magnetosphere emission.  The first,
``Outer Gap'' \citep[OG;][]{chr86a}, model posits emission from vacuum
dipole field lines radiating from the null charge surface to the
light cylinder at radius $R_{\rm LC}$.  In this model, the heuristic
efficiency scaling $\eta \approx (10^{33} {\rm erg\,s^{-1}}/{\dot
E})^{1/2}$ is assumed.  For \psr, the value $\eta = 0.18$ was used
for the gap width $w$ in the OG calculations.  The second model,
``Two Pole Caustic'' \citep[TPC;][]{dr03}, has emission in the vacuum
open zone from the neutron star surface out to 0.75\,$R_{\rm LC}$
from the rotation axis or an altitude of $R_{\rm LC}$, whichever is
lesser.  More recently, numerical models of force-free magnetospheres
have been computed by \citet{bs10} and they have argued that the
pulsar gamma-ray beam can be created in the ``Separatrix Layer''
where the open zone current sheets overlap. The emission in this SL
model is assumed to extend to 1.5\,$R_{\rm LC}$. We assess the match
to the data by plotting a goodness of fit $\chi_3$ surface in the
$\alpha$--$\zeta$ plane \citep[see][]{rw10}. Using the $E > 0.8$\,GeV
light curve, the best-fit OG models (with best $\chi_3= 3.9$ at $\alpha=
133^\circ$, $\zeta=110^\circ$) lie inside the $3\,\sigma$ 1.5\,GHz fit
contours and at the maximum of the combined 1.5\,GHz and 0.8\,GHz fit.
A small second region of adequate but poorer fits ($\chi_3= 7$) is also
present within the $1\,\sigma$ RVM contour at smaller $\alpha= 63^\circ$
($\zeta=60^\circ$).  The SL model has a similar small region of good fits
within the $1\,\sigma$ 1.5\,GHz RVM contour ($\chi_3= 5$ at $\alpha=
50^\circ$, $\zeta=48^\circ$).  The TPC model predicts a large unpulsed
component resulting in much poorer fits (not shown), with a best value
$\chi_3 = 19$ in the RVM-allowed region; this is at $\alpha= 26^\circ$,
$\zeta=25^\circ$, giving a light curve with little pulsed flux.

Figure~\ref{fig:lc} shows the best-fit light curves in the RVM-allowed
region for the OG and SL models, compared with the data. The present
models do not include the full radiation physics, employing simple uniform
emissivity throughout the active zones.  However, it appears that both
high altitude models can produce reasonable matches to the LAT data.
With improved geometrical constraints it may be possible to distinguish
between the OG and SL models.  For the former, the fit phase of the
magnetic axis is at $\phi \approx 0.09$, after the pulse and somewhat
later than the polarization sweep.  The SL model prefers a magnetic
axis much closer to the maximum of the radio pulse.  In both cases,
the lag of the sweep maximum from the pulse peak implies significant
altitude $\sim 0.1$--$0.15\,R_{\rm LC}$ for the radio emission. Such
altitudes are also required to accommodate the relatively wide radio
pulse in the open zone.  Modeling such finite altitude effects may
provide some additional restrictions on the geometry.  We also note that
the OG and SL models predict somewhat different viewing angles $\zeta$;
if an X-ray pulsar wind torus could be measured, the resulting $\zeta$
might be used to select the preferred model.

The classic TPC model fails in these fits at least in part because
it does not have enough emission at high altitudes.  In order to
improve upon this, we also fit the $>0.8$\,GeV profile of \psr\
with light curves predicted from a geometrical high-altitude ``Slot
Gap'' model \citep[SG;][]{mh04a}, an evolution of the low-altitude
SG of \citet{mh03}.  In the SG geometry, the emission originates in
a gap that extends from the neutron star surface at the polar cap
to high altitudes near the light cylinder.  We fit for four model
parameters: $\alpha$, $\zeta$, gap width $w$ \citep[a fraction of the
open volume radius of the polar cap, $r_{\mathrm{ovc}}$, in the open
volume coordinate system described in][]{dh04}, and maximum radius
of emission $r$ (in units of $R_{\mathrm{LC}}$), assuming uniform
emission rate in the corotating frame.  Our resolution is $1^{\circ}$
in $\alpha$ and $\zeta$, $0.01\,r_{\mathrm{ovc}}$ in $0 \leq w \leq
0.3$, and $0.1\,R_{\mathrm{LC}}$ in $0.7\,R_{\mathrm{LC}} \leq r \leq
2.0\,R_{\mathrm{LC}}$ (we assumed a maximum cylindrical emission radius of
$0.95\,R_{\mathrm{LC}}$).  The simulated light curves are fit to the data
via a Markov Chain Monte Carlo (MCMC) routine that explores the parameter
space and records regions of maximum likelihood, as in \citet{vps+03}.

\begin{figure}[t]
\begin{center}
\includegraphics[scale=0.54]{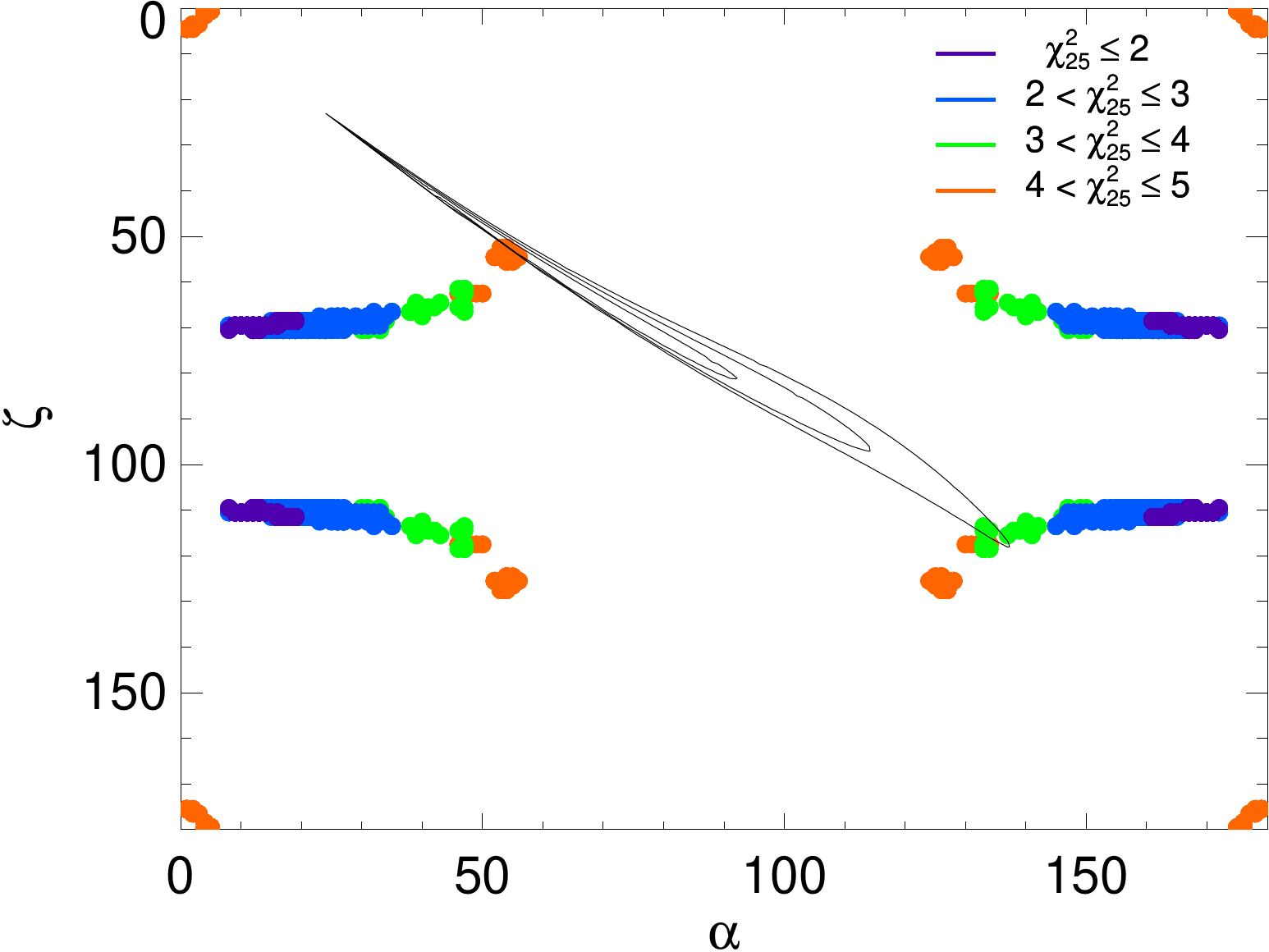}
\end{center}
\vspace{-5mm}
\caption{\label{fig:chi2} Reduced $\chi^2$ values (with 25 degrees of
freedom) in $\alpha$--$\zeta$ space from slot gap gamma-ray light curve
fits, marginalized over $w$, $r$, and $\Delta \phi$, and black curves
from the RVM fit to radio polarization measurements, showing where
these constraints intersect the geometrical light curve parameters
(the overall best-fit light curve lies outside these regions;
see Section~\ref{sec:roger}).  The plotted values of $\chi^2_{25}$
are those included in the Markov chains via the algorithm given in
\citet{vps+03}.  }
\end{figure}

\begin{figure*}[ht]
\begin{center}
\includegraphics[scale=1.04]{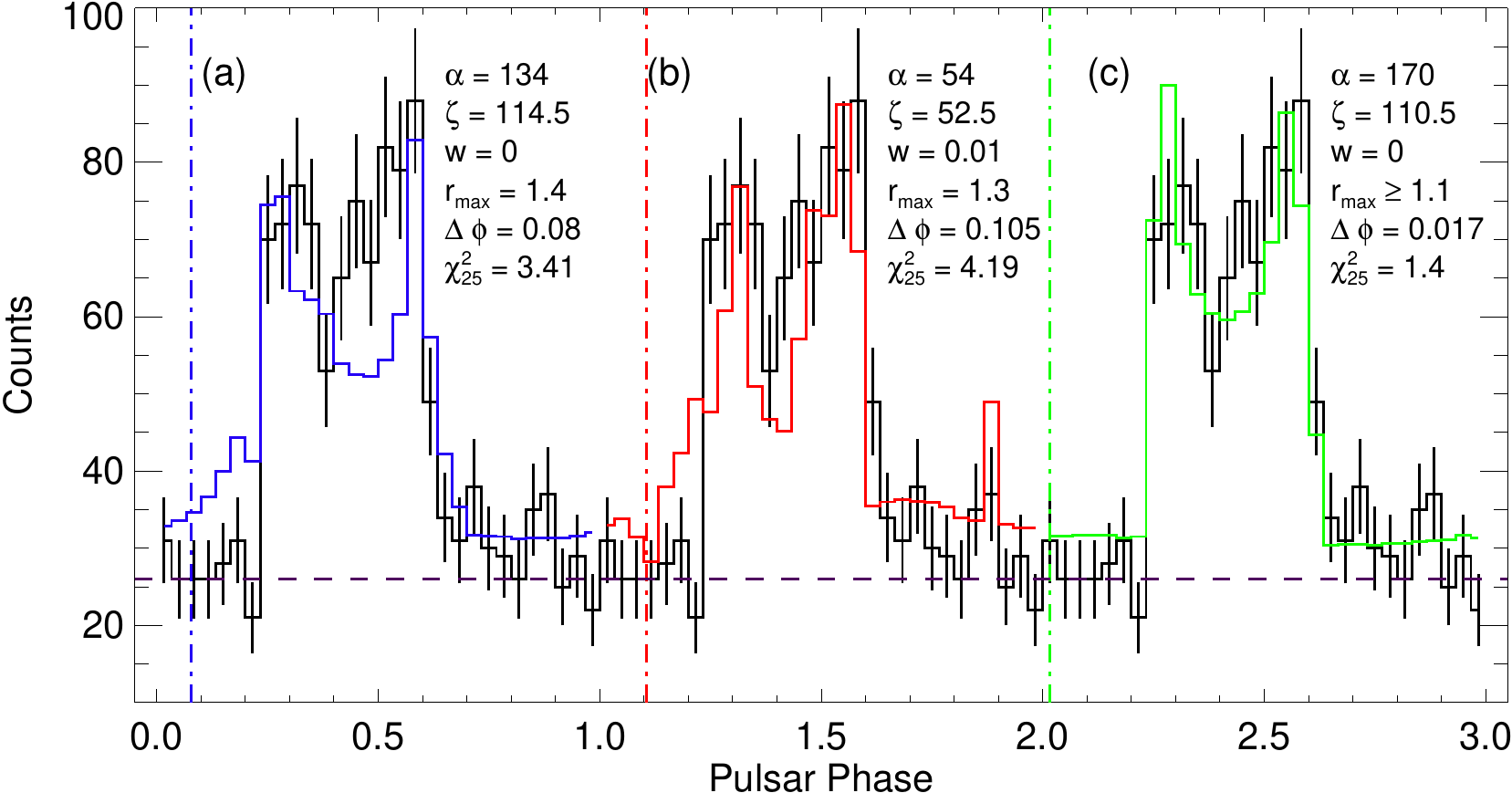}
\end{center}
\caption{\label{fig:3lc}
Slot gap model light curves compared to the data.  (a) The best-fit light
curve (blue) with parameters lying within the polarization contours. The
LAT light curve is shown in black. The vertical dot-dashed line shows
the model location of the magnetic pole, shifted by $\Delta \phi$
from zero. The horizontal dashed line represents the background count
level. (b) Same as (a), for the second-best fit parameters (red). (c)
Same as (a) and (b) for the absolute best-fit light curve (green) found
by the MCMC routine, with no restriction on $\Delta \phi$. Its parameters
do not fall within the radio polarization contours.  }
\end{figure*}

As noted before, we have some independent constraints on the emission
geometry of \psr\ from radio polarization (cf.\ Figure~\ref{fig:pol}).
To make our model fits be consistent with the RVM fit, we allowed the
model light curve to shift with respect to the location of the magnetic
pole by no more than $\Delta \phi = 0.1$ in phase.  Figure~\ref{fig:chi2}
shows the $\alpha$--$\zeta$ space with the polarization contours already
shown in Figure~\ref{fig:chi3} and color ``contours'' of reduced $\chi^2$
from the MCMC likelihood fits. The polarization contours pass through
parameters yielding light curve fits with $3 \leq \chi^2/25 \leq 5$,
the best fit falling within the contours occurring at ($\alpha$, $\zeta$,
$w$, $r$, $\Delta \phi$; $\chi^2/25$) = ($134^{\circ}$, $114.5^{\circ}$,
$0.0\,r_{\mathrm{ovc}}$, $1.4\,R_{\mathrm{LC}}$, 0.08; 3.4).  The second
best fit that falls within the polarization contours is at ($54^{\circ}$,
$52.5^{\circ}$, $0.01\,r_{\mathrm{ovc}}$, $1.3\,R_{\mathrm{LC}}$, 0.105;
4.2).  We include this fit because while it has a larger $\chi^2$, it
falls within a more constrained region of the polarization contours.
The absolute best fit, with or without restricting $\Delta \phi$,
is at ($170^{\circ}$, $110.5^{\circ}$, $0.0\,r_{\mathrm{ovc}}$,
$>1.1\,R_{\mathrm{LC}}$, 0.017; 1.4).  This is inconsistent with standard
radio cone beam emission because the values of $\alpha$ and $\zeta$
are too far apart for lower altitude radio emission to be visible.
All of these light curves are shown in Figure~\ref{fig:3lc}.

Two of the fits include $w=0$, meaning the emission in the model
originates from a single open field line.  Physically, this means the
emission zone width is smaller than the resolution of our simulations,
but not really zero.  All the fits have small widths, requiring higher
current of radiating particles in the gap (either more primaries or
high pair multiplicity) in order to generate the observed luminosity.
It is also likely that the emissivity and width of the emission zone
change azimuthally and with altitude, unlike the constant width of our
simulations \citep[see][]{hm11}.  None of the fits are able to reproduce
well the small middle peak of the observed light curve; it is not yet
clear whether this is a separate feature or part of the second peak.
Also, the background is slightly over-predicted by the SG model,
especially for the parameters consistent with the polarization curves.

None of these (OG, SL, SG) gamma-ray light curve fits are particularly
consistent with our inferences from the radio profile and polarimetric
constraints.  For instance, the latter suggest a nearly orthogonal rotator
(Section~\ref{sec:simon}), for which there are no good gamma-ray light
curve fits.  However, the polarimetric data provide limited constraints,
owing to the small range of pulse longitude for which they are obtained,
and our inferences rely on some assumptions.  As for the gamma-ray fits,
none are perfect, although the OG and SG fits appear slightly better
than the SL one.  Considering their inherent limitations, however,
these fits are encouraging, and further refinements may improve our
understanding of the magnetosphere of middle-aged pulsars like \psr.

For each beaming model, one can extrapolate the average pulsar flux
on the Earth line-of-sight to the isotropic all-sky equivalent.
For the best-fit angles consistent with polarization constraints,
the corrections for \psr\ are $f_\Omega = 0.66$ (OG), 0.64 (SL), and
0.78 (SG; neglecting polarization constraints, $f_{\Omega, \rm SG} =
0.45$).  Using the observed flux and the 8\,kpc DM-estimated distance,
we find that $\eta_{\rm obs} = 4\pi d^2 f_\Omega F_\gamma/{\dot E} =
10\,d_8^2 f_\Omega$.  This indicates that any distance greater that
$\sim 2.5$\,kpc is problematic, unless the emission is highly beamed.
For the OG/SL/SG beaming $f_\Omega \sim 0.7$, we require $d<3$\,kpc.
The preferred distance for the heuristic OG efficiency law ($\eta = 0.18$)
is $\sim 1.3$\,kpc.  Such distance estimates of $\approx 2$\,kpc are
consistent with the discussion at the beginning of Section~\ref{sec:disc},
and strongly suggest that \psr\ is located at roughly $10\times$ the
distance to Geminga ($d=0.16$\,kpc). Improved distance estimates for
middle-aged, apparently efficient pulsars have the potential to constrain
or falsify gamma-ray emission models.

\subsection{Where Have All the (Ordinary Radio) Pulsars Gone?}

In a radio search with the GBT of three unidentified low Galactic latitude
LAT sources, we have discovered \psr, a relatively distant member of the
under-represented group of gamma-ray pulsars that are middle-aged and
have small $\dot E$.  A gamma-ray-only pulsar was subsequently discovered
in one of the two remaining sources, 1FGL~J1746.7--3233 \citep{pga+11}.
The third source could yet be a gamma-ray pulsar, but is unlikely to
be a substantial radio emitter beamed towards the Earth (assuming we
covered its location adequately --- in the case of 1FGL~J1746.7--3233,
the pulsar is $3\farcm8$ offset from our GBT search position; see
Section~\ref{sec:search}): for a fiducial distance of 4\,kpc, our
sensitivity corresponds to a luminosity limit $L_{1.4} \equiv S_{1.4} d^2
= 0.5$\,mJy\,kpc$^2$, which is near the low end of the radio luminosity
distribution for detected radio pulsars \citep[see][which include the
two exceptions with much lower luminosities]{abdo16,cng+09,crr+09}.
It would seem that we were quite successful, with one related radio
discovery among three LAT sources searched.  However, as we argue next,
we don't expect such continued success.

Many more Galactic plane unidentified LAT sources have been searched
for radio pulsars than the three mentioned here.  This work is being
done at the GBT, Parkes, Nan\c{c}ay, GMRT, and Effelsberg telescopes by
members of the \fermi\ pulsar search consortium, and includes searches of
all sources that have pulsar characteristics at all Galactic latitudes.
So far this has resulted in the discovery of more than 30 MSPs, all at
high latitudes \citep[e.g.,][]{cgj+11,kjr+11,rrc+11} --- and \psr.  While
a detailed population analysis will have to wait for the publication of
all the completed searches, the lone example of \psr\ already stands out.
Why have more radio pulsars not been detected among the unidentified
Galactic plane LAT sources?

A point related to this question concerns the very deep radio searches
done of 34 pulsars discovered in blind pulsation searches of gamma rays
\citep[e.g.,][]{abdo2,pga+11,sdz+10}: only four of them were found
to also be radio emitters \citep[see][]{pga+11,rkp+11}.  Three are
exceptionally faint and their detection was possible only following the
gamma-ray discoveries \citep{abdo16,crr+09,pga+11}.  The remaining one,
PSR~J2032+4127 \citep{crr+09}, is like \psr\ in that they are both located
in the Cygnus region and both are relatively bright, being detectable in
1~minute with the GBT at 2\,GHz.  They were not known prior to the launch
of \fermi\ simply because no sensitive surveys had (or have yet fully)
been done of the northern (equatorial) reaches of the Galactic plane,
using the high radio frequencies desired to unveil the Galactic disk.
Conversely, the enormously successful 1.4\,GHz Parkes multibeam survey
\citep[e.g.,][]{mlc+01} {\em did} search the entire Galactic plane at
$260\arcdeg<l<50\arcdeg$ with sensitivity sufficient to detect ordinary
(non-MSP) radio pulsars with $L_{1.4} \ga 0.5$\,mJy\,kpc$^2$ out to $d \ga
2$\,kpc.  A very large fraction of all potentially detectable ordinary
radio pulsars in that area of the sky that could also be plausible
gamma-ray sources were therefore discovered before \fermi\ launch,
especially when adding those discovered in deep pointed observations
of known pulsar wind nebulae \citep[e.g.,][]{cmgl02}.  In time,
several of these pulsars were detected by LAT \citep[e.g.,][]{abdo11}.
Subsequent to the Parkes multibeam survey an even deeper Parkes survey
extended Galactic plane coverage out to $l = 60\arcdeg$ (Camilo et al.,
in preparation), and the on-going PALFA Arecibo survey is doing so out
to $l \approx 75\arcdeg$ \citep{cfl+06}.

In our view, the above suggests that the answer to ``why have more radio
pulsars not been detected among the unidentified Galactic plane LAT
sources?'' is that the vast majority of ordinary radio pulsars accessible
to the current generation of telescopes and located within a few degrees
of the plane at $l \la 75\arcdeg$ that can yield an appreciable gamma-ray
flux at the Earth were discovered a long time ago (it is possible that
in rare but important cases very high $L_{\gamma}$ distant pulsars will
have their radio pulses scattered beyond practical detectability).

Many of the LAT sources that remain unidentified along the Galactic
plane are surely pulsars --- but they may not be detectable as radio
sources, and should be searched anew in gamma rays \citep[some of
these may be in binaries, and inaccessible to current blind searches;
see, e.g.,][]{cck+11}.  \psr\ is bright enough that it would have been
discovered in blind searches of 18 months of LAT data, if the photon
selection criteria had been adjusted to take into account its flat
spectrum (e.g., if only photons with $E\ga0.8$\,GeV had been searched
for pulsations), as we confirmed after discovery.  Spectral analysis
of a region can improve the signal-to-noise ratio by selecting events
based on the probability that they come from the source of interest
\citep[see][]{bel11}, and applying this technique to the Cygnus
region also resulted in the unbiased detection of pulsations from \psr.
For fainter point sources superimposed on the large and uncertain diffuse
Galactic background, spectral analysis and localization are harder
(\fgl\ was only $0\farcm4$ from the actual pulsar position), but these
and other improvements in search techniques are already resulting in the
discovery of many more gamma-ray pulsars \citep[see][]{pga+11,par11}.
As a consequence, the ratio of known gamma-ray-only to gamma-ray-and-radio
ordinary pulsars, which is currently slightly under 1.0, should increase
substantially.

When will we ever learn all that we can (in the \fermi\ era) about
the geometry and emission properties of gamma-ray and radio pulsar
accelerators?  Perhaps when full details emerge from the continuing
collaborative radio and gamma-ray observational and modeling work.
The future is bright (at some wavelength), but requires publication of
all searches, including radio non-detections, and consideration of the
non-detection by LAT of radio pulsars with very large $\dot E$ flux at
the Earth \citep[e.g.,][]{rkc+11}.

\acknowledgements

The GBT is operated by the National Radio Astronomy Observatory, a
facility of the National Science Foundation operated under cooperative
agreement by Associated Universities, Inc.
The \fermi\ LAT Collaboration acknowledges generous ongoing support
from a number of agencies and institutes that have supported both
the development and the operation of the LAT as well as scientific
data analysis.  These include the National Aeronautics and Space
Administration (NASA) and the Department of Energy in the United States,
the Commissariat \`a l'Energie Atomique and the Centre National de la
Recherche Scientifique/Institut National de Physique Nucl\'eaire et de
Physique des Particules in France, the Agenzia Spaziale Italiana and
the Istituto Nazionale di Fisica Nucleare in Italy, the Ministry of
Education, Culture, Sports, Science and Technology (MEXT), High Energy
Accelerator Research Organization (KEK) and Japan Aerospace Exploration
Agency (JAXA) in Japan, and the K.~A.~Wallenberg Foundation, the Swedish
Research Council and the Swedish National Space Board in Sweden.
Support for this work was provided by NASA through Einstein Postdoctoral
Fellowship Award Number PF0-110073 issued by the Chandra X-ray Observatory
Center, which is operated by the Smithsonian Astrophysical Observatory
for and on behalf of NASA under contract NAS8-03060.

{\em Facilities:}  \facility{Fermi (LAT)}, \facility{GBT (GUPPI)},
\facility{Swift (XRT)}


\begin{deluxetable}{ll}
\tablewidth{0.48\linewidth}
\tablecaption{\label{tab:parms} Measured and Derived Parameters for
PSR~J2030+3641 }
\tablecolumns{2}
\tablehead{
\colhead{Parameter} &
\colhead{Value}
}
\startdata
Right ascension, R.A. (J2000.0)\dotfill & $20^{\rm h}30^{\rm m}00\fs261(4)$   \\
Declination, decl. (J2000.0)\dotfill    & $36\arcdeg41'27\farcs15(7)$         \\
Galactic longitude, $l$ (deg)\dotfill   & 76.12                               \\
Galactic latitude, $b$ (deg)\dotfill    & --1.44                              \\
Rotation frequency, $\nu$ ($s^{-1}$)\dotfill & 4.99678736138(4)               \\
Frequency derivative, $\dot \nu$ ($s^{-2}$)\dotfill 
                                        & $-1.62294(5)\times10^{-13}$         \\
Frequency second derivative, $\ddot \nu$ ($s^{-3}$)\dotfill 
                                        & $2.2(6)\times10^{-24}$              \\
Epoch of frequency (MJD)\dotfill        & 55400.0                             \\
Dispersion measure, DM (\pccc)\dotfill  & 246.0(7)                            \\
Data span (MJD)\dotfill                 & 54700--55671                        \\
RMS timing residual (ms)\dotfill        & 0.4                                 \\
Spin-down luminosity, $\dot E$ (${\rm erg\,s^{-1}}$)\dotfill
                                                       & $3.2\times10^{34}$   \\
Characteristic age, $\tau_c$ (yr)\dotfill              & $4.9\times10^5$      \\
Surface dipole magnetic field strength (Gauss)\dotfill & $1.2\times10^{12}$   \\
Flux density at 2\,GHz, $S_2$ (mJy)\dotfill             & 0.09                \\
Flux density at 1.5\,GHz, $S_{1.5}$ (mJy)\dotfill       & 0.15                \\
Flux density at 0.8\,GHz, $S_{0.8}$ (mJy)\dotfill       & 0.40                \\
Rotation measure, RM (rad\,m$^{-2}$)\dotfill            & $+514\pm1$          \\
Radio--gamma-ray profile offset, $\delta$ ($P$)\dotfill & $0.26\pm0.02$       \\
Gamma-ray profile peak-to-peak separation, $\Delta$ ($P$)\dotfill
                                                 & $0.30^{+0.03}_{-0.08}$     \\
Gamma-ray ($>0.1$\,GeV) photon index, $\Gamma$\dotfill
                                                 & $1.1\pm0.2^{+0.2}_{-0.3}$  \\
Gamma-ray cut-off energy, $E_c$ (GeV)\dotfill    & $2.0\pm0.3^{+0.3}_{-0.4}$  \\
Photon flux ($>0.1$\,GeV) ($10^{-8}\,{\rm cm^{-2}\,s^{-1}}$)\dotfill
                                                 & $3.6\pm0.5\pm0.8$          \\
Energy flux ($>0.1$\,GeV), $F_{\gamma}$
($10^{-11}\,{\rm erg\,cm^{-2}\,s^{-1}}$)\dotfill & $4.2\pm0.3\pm0.5$          \\
DM-derived distance (kpc)\dotfill                        &  8                 \\
Most likely distance\tablenotemark{a}, $d$ (kpc)\dotfill &  1.5--3      \\[-5pt]
\enddata
\tablecomments{Numbers in parentheses represent the nominal $1\,\sigma$
TEMPO2 timing uncertainties on the last digits quoted.  For gamma-ray
parameters, the first uncertainty is statistical and the second accounts
for systematics (see Section~\ref{sec:spect}). }
\tablenotetext{a}{See discussion at the beginning of
Section~\ref{sec:disc} and at the end of Section~\ref{sec:roger}. }
\end{deluxetable}


\begin{thebibliography}{54}
\expandafter\ifx\csname natexlab\endcsname\relax\def\natexlab#1{#1}\fi

\bibitem[{{Abdo} {et~al.}(2007){Abdo}, {Allen}, {Berley}, {Casanova}, {Chen},
  {Coyne}, {Dingus}, {Ellsworth}, {Fleysher}, {Fleysher}, {Gonzalez},
  {Goodman}, {Hays}, {Hoffman}, {Hopper}, {H{\"u}ntemeyer}, {Kolterman},
  {Lansdell}, {Linnemann}, {McEnery}, {Mincer}, {Nemethy}, {Noyes}, {Ryan},
  {Saz Parkinson}, {Shoup}, {Sinnis}, {Smith}, {Sullivan}, {Vasileiou},
  {Walker}, {Williams}, {Xu}, \& {Yodh}}]{aab+07}
{Abdo}, A.~A., {et~al.} 2007, \apjl, 664, L91

\bibitem[{{Abdo} {et~al.}(2009{\natexlab{a}}){Abdo}, {Ackermann}, {Ajello},
  {Atwood}, {Axelsson}, {Baldini}, {Ballet}, {Barbiellini}, {Baring},
  {Bastieri}, {Baughman}, {Bechtol}, {Bellazzini}, {Berenji}, {Bignami},
  {Blandford}, {Bloom}, {Bonamente}, {Borgland}, {Bregeon}, {Brez}, {Brigida},
  {Bruel}, {Burnett}, {Caliandro}, {Cameron}, {Camilo}, {Caraveo}, {Carlson},
  {Casandjian}, {Cecchi}, {{\c C}elik}, {Charles}, {Chekhtman}, {Cheung},
  {Chiang}, {Ciprini}, {Claus}, {Cognard}, {Cohen-Tanugi}, {Cominsky},
  {Conrad}, {Corbet}, {Cutini}, {Dermer}, {Desvignes}, {de Angelis}, {de Luca},
  {de Palma}, {Digel}, {Dormody}, {do Couto e Silva}, {Drell}, {Dubois},
  {Dumora}, {Edmonds}, {Farnier}, {Favuzzi}, {Fegan}, {Focke}, {Frailis},
  {Freire}, {Fukazawa}, {Funk}, {Fusco}, {Gargano}, {Gasparrini}, {Gehrels},
  {Germani}, {Giebels}, {Giglietto}, {Giordano}, {Glanzman}, {Godfrey},
  {Grenier}, {Grondin}, {Grove}, {Guillemot}, {Guiriec}, {Hanabata}, {Harding},
  {Hayashida}, {Hays}, {Hobbs}, {Hughes}, {J{\'o}hannesson}, {Johnson},
  {Johnson}, {Johnson}, {Johnson}, {Johnston}, {Kamae}, {Katagiri}, {Kataoka},
  {Kawai}, {Kerr}, {Kn{\"o}dlseder}, {Kocian}, {Kramer}, {Kuss}, {Lande},
  {Latronico}, {Lemoine-Goumard}, {Longo}, {Loparco}, {Lott}, {Lovellette},
  {Lubrano}, {Madejski}, {Makeev}, {Manchester}, {Marelli}, {Mazziotta},
  {McConville}, {McEnery}, {McLaughlin}, {Meurer}, {Michelson}, {Mitthumsiri},
  {Mizuno}, {Moiseev}, {Monte}, {Monzani}, {Morselli}, {Moskalenko}, {Murgia},
  {Nolan}, {Norris}, {Nuss}, {Ohsugi}, {Omodei}, {Orlando}, {Ormes}, {Paneque},
  {Panetta}, {Parent}, {Pelassa}, {Pepe}, {Pesce-Rollins}, {Piron}, {Porter},
  {Rain{\`o}}, {Rando}, {Ransom}, {Ray}, {Razzano}, {Rea}, {Reimer}, {Reimer},
  {Reposeur}, {Ritz}, {Rochester}, {Rodriguez}, {Romani}, {Roth}, {Ryde},
  {Sadrozinski}, {Sanchez}, {Sander}, {Saz Parkinson}, {Scargle}, {Schalk},
  {Sgr{\`o}}, {Siskind}, {Smith}, {Smith}, {Spandre}, {Spinelli}, {Stappers},
  {Starck}, {Striani}, {Strickman}, {Suson}, {Tajima}, {Takahashi}, {Tanaka},
  {Thayer}, {Thayer}, {Theureau}, {Thompson}, {Thorsett}, {Tibaldo}, {Torres},
  {Tosti}, {Tramacere}, {Uchiyama}, {Usher}, {Van Etten}, {Vasileiou},
  {Venter}, {Vilchez}, {Vitale}, {Waite}, {Wallace}, {Wang}, {Watters}, {Webb},
  {Weltevrede}, {Winer}, {Wood}, {Ylinen}, \& {Ziegler}}]{abdo10}
---. 2009{\natexlab{a}}, Science, 325, 848

\bibitem[{{Abdo} {et~al.}(2009{\natexlab{b}}){Abdo}, {Ackermann}, {Ajello},
  {Anderson}, {Atwood}, {Axelsson}, {Baldini}, {Ballet}, {Barbiellini},
  {Baring}, {Bastieri}, {Baughman}, {Bechtol}, {Bellazzini}, {Berenji},
  {Bignami}, {Blandford}, {Bloom}, {Bonamente}, {Borgland}, {Bregeon}, {Brez},
  {Brigida}, {Bruel}, {Burnett}, {Caliandro}, {Cameron}, {Caraveo},
  {Casandjian}, {Cecchi}, {{\c C}elik}, {Chekhtman}, {Cheung}, {Chiang},
  {Ciprini}, {Claus}, {Cohen-Tanugi}, {Conrad}, {Cutini}, {Dermer}, {de
  Angelis}, {de Luca}, {de Palma}, {Digel}, {Dormody}, {do Couto e Silva},
  {Drell}, {Dubois}, {Dumora}, {Farnier}, {Favuzzi}, {Fegan}, {Fukazawa},
  {Funk}, {Fusco}, {Gargano}, {Gasparrini}, {Gehrels}, {Germani}, {Giebels},
  {Giglietto}, {Giommi}, {Giordano}, {Glanzman}, {Godfrey}, {Grenier},
  {Grondin}, {Grove}, {Guillemot}, {Guiriec}, {Gwon}, {Hanabata}, {Harding},
  {Hayashida}, {Hays}, {Hughes}, {J{\'o}hannesson}, {Johnson}, {Johnson},
  {Johnson}, {Kamae}, {Katagiri}, {Kataoka}, {Kawai}, {Kerr}, {Kn{\"o}dlseder},
  {Kocian}, {Kuss}, {Lande}, {Latronico}, {Lemoine-Goumard}, {Longo},
  {Loparco}, {Lott}, {Lovellette}, {Lubrano}, {Madejski}, {Makeev}, {Marelli},
  {Mazziotta}, {McConville}, {McEnery}, {Meurer}, {Michelson}, {Mitthumsiri},
  {Mizuno}, {Monte}, {Monzani}, {Morselli}, {Moskalenko}, {Murgia}, {Nolan},
  {Norris}, {Nuss}, {Ohsugi}, {Omodei}, {Orlando}, {Ormes}, {Paneque},
  {Parent}, {Pelassa}, {Pepe}, {Pesce-Rollins}, {Pierbattista}, {Piron},
  {Porter}, {Primack}, {Rain{\`o}}, {Rando}, {Ray}, {Razzano}, {Rea}, {Reimer},
  {Reimer}, {Reposeur}, {Ritz}, {Rochester}, {Rodriguez}, {Romani}, {Ryde},
  {Sadrozinski}, {Sanchez}, {Sander}, {Parkinson}, {Scargle}, {Sgr{\`o}},
  {Siskind}, {Smith}, {Smith}, {Spandre}, {Spinelli}, {Starck}, {Strickman},
  {Suson}, {Tajima}, {Takahashi}, {Takahashi}, {Tanaka}, {Thayer}, {Thompson},
  {Tibaldo}, {Tibolla}, {Torres}, {Tosti}, {Tramacere}, {Uchiyama}, {Usher},
  {Van Etten}, {Vasileiou}, {Vilchez}, {Vitale}, {Waite}, {Wang}, {Watters},
  {Winer}, {Wolff}, {Wood}, {Ylinen}, \& {Ziegler}}]{abdo2}
---. 2009{\natexlab{b}}, Science, 325, 840

\bibitem[{{Abdo} {et~al.}(2009{\natexlab{c}}){Abdo}, {Ackermann}, {Atwood},
  {Bagagli}, {Baldini}, {Ballet}, {Band}, {Barbiellini}, {Baring}, {Bartelt},
  \& et~al.}]{abdo2009}
---. 2009{\natexlab{c}}, \apj, 696, 1084

\bibitem[{{Abdo} {et~al.}(2009{\natexlab{d}}){Abdo}, {Allen}, {Aune}, {Berley},
  {Chen}, {Christopher}, {DeYoung}, {Dingus}, {Ellsworth}, {Gonzalez},
  {Goodman}, {Hays}, {Hoffman}, {H{\"u}ntemeyer}, {Kolterman}, {Linnemann},
  {McEnery}, {Morgan}, {Mincer}, {Nemethy}, {Pretz}, {Ryan}, {Saz Parkinson},
  {Shoup}, {Sinnis}, {Smith}, {Vasileiou}, {Walker}, {Williams}, \&
  {Yodh}}]{milagro09}
---. 2009{\natexlab{d}}, \apjl, 700, L127

\bibitem[{{Abdo} {et~al.}(2009{\natexlab{e}}){Abdo}, {Ackermann}, {Ajello},
  {Atwood}, {Baldini}, {Ballet}, {Barbiellini}, {Bastieri}, {Battelino},
  {Baughman}, {Bechtol}, {Bellazzini}, {Berenji}, {Bloom}, {Bogaert},
  {Borgland}, {Bregeon}, {Brez}, {Brigida}, {Bruel}, {Burnett}, {Caliandro},
  {Cameron}, {Camilo}, {Caraveo}, {Casandjian}, {Cecchi}, {Charles},
  {Chekhtman}, {Chen}, {Cheung}, {Chiang}, {Ciprini}, {Cognard},
  {Cohen-Tanugi}, {Cominsky}, {Conrad}, {Cutini}, {Demorest}, {Dermer}, {de
  Angelis}, {de Luca}, {de Palma}, {Digel}, {Dormody}, {do Couto e Silva},
  {Drell}, {Dubois}, {Dumora}, {Espinoza}, {Farnier}, {Favuzzi}, {Focke},
  {Frailis}, {Freire}, {Fukazawa}, {Funk}, {Fusco}, {Gargano}, {Gasparrini},
  {Gehrels}, {Germani}, {Giebels}, {Giglietto}, {Giordano}, {Glanzman},
  {Godfrey}, {Grenier}, {Grondin}, {Grove}, {Guillemot}, {Guiriec}, {Hanabata},
  {Harding}, {Hayashida}, {Hays}, {Hughes}, {J{\'o}hannesson}, {Johnson},
  {Johnson}, {Johnson}, {Johnson}, {Johnston}, {Kamae}, {Katagiri}, {Kataoka},
  {Kawai}, {Kerr}, {Kiziltan}, {Kn{\"o}dlseder}, {Komin}, {Kramer}, {Kuehn},
  {Kuss}, {Lande}, {Latronico}, {Lee}, {Lemoine-Goumard}, {Longo}, {Loparco},
  {Lott}, {Lovellette}, {Lubrano}, {Lyne}, {Makeev}, {Manchester}, {Marelli},
  {Mazziotta}, {McConville}, {McEnery}, {McLaughlin}, {Meurer}, {Michelson},
  {Mitthumsiri}, {Mizuno}, {Moiseev}, {Monte}, {Monzani}, {Morselli},
  {Moskalenko}, {Murgia}, {Nolan}, {Noutsos}, {Nuss}, {Ohsugi}, {Omodei},
  {Orlando}, {Ormes}, {Ozaki}, {Paneque}, {Panetta}, {Parent}, {Pepe},
  {Pesce-Rollins}, {Piron}, {Porter}, {Rain{\`o}}, {Rando}, {Ransom},
  {Razzano}, {Reimer}, {Reimer}, {Reposeur}, {Ritz}, {Rochester}, {Rodriguez},
  {Romani}, {Ryde}, {Sadrozinski}, {Sanchez}, {Parkinson}, {Sgr{\`o}},
  {Sierpowska-Bartosik}, {Siskind}, {Smith}, {Smith}, {Spandre}, {Spinelli},
  {Stappers}, {Starck}, {Strickman}, {Suson}, {Tajima}, {Takahashi},
  {Takahashi}, {Tanaka}, {Thayer}, {Thayer}, {Theureau}, {Thompson},
  {Thorsett}, {Tibaldo}, {Torres}, {Tosti}, {Tramacere}, {Uchiyama}, {Usher},
  {Van Etten}, {Vilchez}, {Vitale}, {Waite}, {Wallace}, {Watters},
  {Weltevrede}, {Wood}, {Ylinen}, \& {Ziegler}}]{abdo15}
---. 2009{\natexlab{e}}, \apj, 700, 1059

\bibitem[{{Abdo} {et~al.}(2010{\natexlab{a}}){Abdo}, {Ackermann}, {Ajello},
  {Allafort}, {Baldini}, {Ballet}, {Barbiellini}, {Bastieri}, {Bechtol},
  {Bellazzini}, {Berenji}, {Blandford}, {Bloom}, {Bonamente}, {Borgland},
  {Bouvier}, {Bregeon}, {Brez}, {Brigida}, {Bruel}, {Burnett}, {Buson},
  {Caliandro}, {Cameron}, {Camilo}, {Caraveo}, {Carrigan}, {Casandjian},
  {Cecchi}, {{\c C}elik}, {Chekhtman}, {Cheung}, {Chiang}, {Ciprini}, {Claus},
  {Cognard}, {Cohen-Tanugi}, {Conrad}, {Corbet}, {DeCesar}, {Dermer},
  {Desvignes}, {de Angelis}, {de Palma}, {Digel}, {Dormody}, {Silva}, {Drell},
  {Dubois}, {Dumora}, {Espinoza}, {Farnier}, {Favuzzi}, {Fegan}, {Focke},
  {Frailis}, {Freire}, {Fukazawa}, {Funk}, {Fusco}, {Gargano}, {Gasparrini},
  {Gehrels}, {Germani}, {Giavitto}, {Giglietto}, {Giordano}, {Glanzman},
  {Godfrey}, {Grenier}, {Grondin}, {Grove}, {Guillemot}, {Guiriec}, {Hadasch},
  {Harding}, {Hays}, {Hobbs}, {Horan}, {Hughes}, {J{\'o}hannesson}, {Johnson},
  {Johnson}, {Johnson}, {Johnston}, {Kamae}, {Katagiri}, {Kataoka}, {Kawai},
  {Kerr}, {Kn{\"o}dlseder}, {Kramer}, {Kuss}, {Lande}, {Latronico},
  {Lemoine-Goumard}, {Llena Garde}, {Longo}, {Loparco}, {Lott}, {Lovellette},
  {Lubrano}, {Lyne}, {Makeev}, {Manchester}, {Marelli}, {Mazziotta},
  {McConville}, {McEnery}, {McGlynn}, {Meurer}, {Michelson}, {Mitthumsiri},
  {Mizuno}, {Moiseev}, {Monte}, {Monzani}, {Morselli}, {Moskalenko}, {Murgia},
  {Nolan}, {Norris}, {Noutsos}, {Nuss}, {Ohsugi}, {Omodei}, {Orlando}, {Ormes},
  {Ozaki}, {Paneque}, {Panetta}, {Parent}, {Pelassa}, {Pepe}, {Pesce-Rollins},
  {Pierbattista}, {Piron}, {Porter}, {Rain{\`o}}, {Rando}, {Ransom}, {Razzano},
  {Reimer}, {Reimer}, {Reposeur}, {Ripken}, {Ritz}, {Rochester}, {Rodriguez},
  {Romani}, {Roth}, {Ryde}, {Sadrozinski}, {Sander}, {Saz Parkinson},
  {Scargle}, {Sgr{\`o}}, {Siskind}, {Smith}, {Smith}, {Spandre}, {Spinelli},
  {Stappers}, {Starck}, {Strickman}, {Suson}, {Takahashi}, {Tanaka}, {Thayer},
  {Thayer}, {Theureau}, {Thompson}, {Thorsett}, {Tibaldo}, {Torres}, {Tosti},
  {Tramacere}, {Usher}, {Van Etten}, {Vasileiou}, {Venter}, {Vilchez},
  {Vitale}, {Waite}, {Wallace}, {Wang}, {Weltevrede}, {Winer}, {Wood},
  {Ylinen}, \& {Ziegler}}]{abdo12}
---. 2010{\natexlab{a}}, \apj, 712, 957

\bibitem[{{Abdo} {et~al.}(2010{\natexlab{b}}){Abdo}, {Ackermann}, {Ajello},
  {Allafort}, {Antolini}, {Atwood}, {Axelsson}, {Baldini}, {Ballet},
  {Barbiellini}, \& et~al.}]{abdo14}
---. 2010{\natexlab{b}}, \apjs, 188, 405

\bibitem[{{Abdo} {et~al.}(2010{\natexlab{c}}){Abdo}, {Ackermann}, {Ajello},
  {Baldini}, {Ballet}, {Barbiellini}, {Bastieri}, {Baughman}, {Bechtol},
  {Bellazzini}, {Berenji}, {Blandford}, {Bloom}, {Bonamente}, {Borgland},
  {Bregeon}, {Brez}, {Brigida}, {Bruel}, {Burnett}, {Buson}, {Caliandro},
  {Cameron}, {Camilo}, {Caraveo}, {Casandjian}, {Cecchi}, {{\c C}elik},
  {Chekhtman}, {Cheung}, {Chiang}, {Ciprini}, {Claus}, {Cognard},
  {Cohen-Tanugi}, {Cominsky}, {Conrad}, {Cutini}, {de Angelis}, {de Palma},
  {Digel}, {Dingus}, {Dormody}, {Silva}, {Drell}, {Dubois}, {Dumora},
  {Farnier}, {Favuzzi}, {Fegan}, {Focke}, {Fortin}, {Frailis}, {Freire},
  {Fukazawa}, {Funk}, {Fusco}, {Gargano}, {Gasparrini}, {Gehrels}, {Germani},
  {Giavitto}, {Giebels}, {Giglietto}, {Giordano}, {Glanzman}, {Godfrey},
  {Grenier}, {Grondin}, {Grove}, {Guillemot}, {Guiriec}, {Hanabata}, {Harding},
  {Hays}, {Hughes}, {Jackson}, {J{\'o}hannesson}, {Johnson}, {Johnson},
  {Johnson}, {Johnston}, {Kamae}, {Katagiri}, {Kataoka}, {Kawai}, {Kerr},
  {Kn{\"o}dlseder}, {Kocian}, {Kuss}, {Lande}, {Latronico}, {Lemoine-Goumard},
  {Longo}, {Loparco}, {Lott}, {Lovellette}, {Lubrano}, {Makeev}, {Marelli},
  {Mazziotta}, {McEnery}, {Meurer}, {Michelson}, {Mitthumsiri}, {Mizuno},
  {Moiseev}, {Monte}, {Monzani}, {Morselli}, {Moskalenko}, {Murgia}, {Nolan},
  {Norris}, {Nuss}, {Ohsugi}, {Omodei}, {Orlando}, {Ormes}, {Paneque},
  {Parent}, {Pelassa}, {Pepe}, {Pesce-Rollins}, {Piron}, {Porter}, {Rain{\`o}},
  {Rando}, {Ray}, {Razzano}, {Reimer}, {Reimer}, {Reposeur}, {Ritz}, {Roberts},
  {Rochester}, {Rodriguez}, {Ro'mani}, {Roth}, {Ryde}, {Sadrozinski},
  {Sanchez}, {Sander}, {Saz Parkinson}, {Scargle}, {Sgr{\`o}}, {Siskind},
  {Smith}, {Smith}, {Spandre}, {Spinelli}, {Strickman}, {Suson}, {Tajima},
  {Takahashi}, {Tanaka}, {Thayer}, {Thayer}, {Theureau}, {Thompson}, {Tibaldo},
  {Tibolla}, {Torres}, {Tosti}, {Tramacere}, {Uchiyama}, {Usher}, {Van Etten},
  {Vasileiou}, {Venter}, {Vilchez}, {Vitale}, {Waite}, {Wang}, {Watters},
  {Winer}, {Wolff}, {Wood}, {Ylinen}, \& {Ziegler}}]{abdo16}
---. 2010{\natexlab{c}}, \apj, 711, 64

\bibitem[{{Abdo} {et~al.}(2010{\natexlab{d}}){Abdo}, {Ackermann}, {Ajello},
  {Atwood}, {Axelsson}, {Baldini}, {Ballet}, {Barbiellini}, {Baring},
  {Bastieri}, \& et~al.}]{abdo11}
---. 2010{\natexlab{d}}, \apjs, 187, 460

\bibitem[{{Abdo} {et~al.}(2011{\natexlab{a}}){Abdo}, {Ackermann}, {Ajello},
  {Atwood}, {Axelsson}, {Baldini}, {Ballet}, {Barbiellini}, {Baring},
  {Bastieri}, \& et~al.}]{abdo17}
---. 2011{\natexlab{a}}, \apjs, submitted (arXiv/1108.1435)

\bibitem[{{Abdo} {et~al.}(2011{\natexlab{b}}){Abdo}, {Wood}, {DeCesar},
  {Gargano}, {Giordano}, {Ray}, {Parent}, {Harding}, {Miller}, {Wood}, \&
  {Wolff}}]{awd+11}
---. 2011{\natexlab{b}}, \apj, submitted (arXiv/1107.4151)

\bibitem[{{Arons}(1996)}]{aro96b}
{Arons}, J. 1996, \aaps, 120, C49

\bibitem[{{Atwood} {et~al.}(2009){Atwood}, {Abdo}, {Ackermann}, {Althouse},
  {Anderson}, {Axelsson}, {Baldini}, {Ballet}, {Band}, {Barbiellini}, \&
  et~al.}]{aaa+09}
{Atwood}, W.~B., {et~al.} 2009, \apj, 697, 1071

\bibitem[{{Bai} \& {Spitkovsky}(2010)}]{bs10}
{Bai}, X., \& {Spitkovsky}, A. 2010, \apj, 715, 1282

\bibitem[{Belfiore(2011)}]{bel11}
Belfiore, A. 2011, in Radio Pulsars: an Astrophysical Key to Unlock the Secrets
  of the Universe, ed. M.~Burgay, N.~D'Amico, P.~Esposito, A.~Pellizzoni, \&
  A.~Possenti, Vol. 1357 (AIP), 44

\bibitem[{Blaskiewicz {et~al.}(1991)Blaskiewicz, Cordes, \& Wasserman}]{bcw91}
Blaskiewicz, M., Cordes, J.~M., \& Wasserman, I. 1991, ApJ, 370, 643

\bibitem[{{Burrows} {et~al.}(2005){Burrows}, {Hill}, {Nousek}, {Kennea},
  {Wells}, {Osborne}, {Abbey}, {Beardmore}, {Mukerjee}, {Short}, {Chincarini},
  {Campana}, {Citterio}, {Moretti}, {Pagani}, {Tagliaferri}, {Giommi},
  {Capalbi}, {Tamburelli}, {Angelini}, {Cusumano}, {Br{\"a}uninger}, {Burkert},
  \& {Hartner}}]{bhn+05}
{Burrows}, D.~N., {et~al.} 2005, \ssr, 120, 165

\bibitem[{Camilo {et~al.}(2002)Camilo, Manchester, Gaensler, \&
  Lorimer}]{cmgl02}
Camilo, F., Manchester, R.~N., Gaensler, B.~M., \& Lorimer, D.~R. 2002, ApJ,
  579, L25

\bibitem[{{Camilo} {et~al.}(2009{\natexlab{a}}){Camilo}, {Ng}, {Gaensler},
  {Ransom}, {Chatterjee}, {Reynolds}, \& {Sarkissian}}]{cng+09}
{Camilo}, F., {Ng}, C.-Y., {Gaensler}, B.~M., {Ransom}, S.~M., {Chatterjee},
  S., {Reynolds}, J., \& {Sarkissian}, J. 2009{\natexlab{a}}, \apjl, 703, L55

\bibitem[{{Camilo} {et~al.}(2009{\natexlab{b}}){Camilo}, {Ray}, {Ransom},
  {Burgay}, {Johnson}, {Kerr}, {Gotthelf}, {Halpern}, {Reynolds}, {Romani},
  {Demorest}, {Johnston}, {van Straten}, {Saz Parkinson}, {Ziegler}, {Dormody},
  {Thompson}, {Smith}, {Harding}, {Abdo}, {Crawford}, {Freire}, {Keith},
  {Kramer}, {Roberts}, {Weltevrede}, \& {Wood}}]{crr+09}
{Camilo}, F., {et~al.} 2009{\natexlab{b}}, \apj, 705, 1

\bibitem[{Cheng {et~al.}(1986)Cheng, Ho, \& Ruderman}]{chr86a}
Cheng, K.~S., Ho, C., \& Ruderman, M. 1986, ApJ, 300, 500

\bibitem[{{Cognard} {et~al.}(2011){Cognard}, {Guillemot}, {Johnson}, {Smith},
  {Venter}, {Harding}, {Wolff}, {Cheung}, {Donato}, {Abdo}, {Ballet}, {Camilo},
  {Desvignes}, {Dumora}, {Ferrara}, {Freire}, {Grove}, {Johnston}, {Keith},
  {Kramer}, {Lyne}, {Michelson}, {Parent}, {Ransom}, {Ray}, {Romani}, {Saz
  Parkinson}, {Stappers}, {Theureau}, {Thompson}, {Weltevrede}, \&
  {Wood}}]{cgj+11}
{Cognard}, I., {et~al.} 2011, \apj, 732, 47

\bibitem[{Corbet {et~al.}(2011)Corbet, Cheung, Kerr, Dubois, Donato, Caliandro,
  Coes, Edwards, Filipovic, Payne, \& Stevens}]{cck+11}
Corbet, R.~H.~D., {et~al.} 2011, 1FGL J1018.6-5856: a new gamma-ray binary,
  {ATEL 3221}

\bibitem[{{Cordes} \& {Lazio}(2002)}]{cl02}
{Cordes}, J.~M., \& {Lazio}, T.~J.~W. 2002, preprint (arXiv:astro-ph/0207156)

\bibitem[{{Cordes} {et~al.}(2006){Cordes}, {Freire}, {Lorimer}, {Camilo},
  {Champion}, {Nice}, {Ramachandran}, {Hessels}, {Vlemmings}, {van Leeuwen},
  {Ransom}, {Bhat}, {Arzoumanian}, {McLaughlin}, {Kaspi}, {Kasian}, {Deneva},
  {Reid}, {Chatterjee}, {Han}, {Backer}, {Stairs}, {Deshpande}, \&
  {Faucher-Gigu{\`e}re}}]{cfl+06}
{Cordes}, J.~M., {et~al.} 2006, ApJ, 637, 446

\bibitem[{{de Jager} \& {B{\"u}sching}(2010)}]{db10}
{de Jager}, O.~C., \& {B{\"u}sching}, I. 2010, \aap, 517, L9

\bibitem[{{Dyks} \& {Harding}(2004)}]{dh04}
{Dyks}, J., \& {Harding}, A.~K. 2004, ApJ, 614, 869

\bibitem[{{Dyks} \& {Rudak}(2003)}]{dr03}
{Dyks}, J., \& {Rudak}, B. 2003, ApJ, 598, 1201

\bibitem[{{Everett} \& {Weisberg}(2001)}]{ew01}
{Everett}, J.~E., \& {Weisberg}, J.~M. 2001, ApJ, 553, 341

\bibitem[{{Gehrels} {et~al.}(2004){Gehrels}, {Chincarini}, {Giommi}, {Mason},
  {Nousek}, {Wells}, {White}, {Barthelmy}, {Burrows}, {Cominsky}, {Hurley},
  {Marshall}, {M{\'e}sz{\'a}ros}, {Roming}, {Angelini}, {Barbier}, {Belloni},
  {Campana}, {Caraveo}, {Chester}, {Citterio}, {Cline}, {Cropper}, {Cummings},
  {Dean}, {Feigelson}, {Fenimore}, {Frail}, {Fruchter}, {Garmire}, {Gendreau},
  {Ghisellini}, {Greiner}, {Hill}, {Hunsberger}, {Krimm}, {Kulkarni}, {Kumar},
  {Lebrun}, {Lloyd-Ronning}, {Markwardt}, {Mattson}, {Mushotzky}, {Norris},
  {Osborne}, {Paczynski}, {Palmer}, {Park}, {Parsons}, {Paul}, {Rees},
  {Reynolds}, {Rhoads}, {Sasseen}, {Schaefer}, {Short}, {Smale}, {Smith},
  {Stella}, {Tagliaferri}, {Takahashi}, {Tashiro}, {Townsley}, {Tueller},
  {Turner}, {Vietri}, {Voges}, {Ward}, {Willingale}, {Zerbi}, \&
  {Zhang}}]{gcg+04}
{Gehrels}, N., {et~al.} 2004, \apj, 611, 1005

\bibitem[{Han {et~al.}(2006)Han, Manchester, Lyne, Qiao, \& van
  Straten}]{hml+06}
Han, J.~L., Manchester, R.~N., Lyne, A.~G., Qiao, G.~J., \& van Straten, W.
  2006, ApJ, 642, 868

\bibitem[{{Harding} \& {Muslimov}(2011)}]{hm11}
{Harding}, A.~K., \& {Muslimov}, A.~G. 2011, \apjl, 726, L10

\bibitem[{{Hobbs} {et~al.}(2006){Hobbs}, {Edwards}, \& {Manchester}}]{hem06}
{Hobbs}, G.~B., {Edwards}, R.~T., \& {Manchester}, R.~N. 2006, MNRAS, 369, 655

\bibitem[{{Hotan} {et~al.}(2004){Hotan}, {van Straten}, \&
  {Manchester}}]{hvm04}
{Hotan}, A.~W., {van Straten}, W., \& {Manchester}, R.~N. 2004, PASA, 21, 302

\bibitem[{{Johnston} \& {Weisberg}(2006)}]{jw06}
{Johnston}, S., \& {Weisberg}, J.~M. 2006, MNRAS, 368, 1856

\bibitem[{{Kargaltsev} \& {Pavlov}(2008)}]{kp08}
{Kargaltsev}, O., \& {Pavlov}, G.~G. 2008, in American Institute of Physics
  Conference Series, Vol. 983, 40 Years of Pulsars: Millisecond Pulsars,
  Magnetars and More, ed. {C.~Bassa, Z.~Wang, A.~Cumming, \& V.~M.~Kaspi},
  171--185

\bibitem[{{Keith} {et~al.}(2011){Keith}, {Johnston}, {Ray}, {Ferrara}, {Saz
  Parkinson}, {{\c C}elik}, {Belfiore}, {Donato}, {Cheung}, {Abdo}, {Camilo},
  {Freire}, {Guillemot}, {Harding}, {Kramer}, {Michelson}, {Ransom}, {Romani},
  {Smith}, {Thompson}, {Weltevrede}, \& {Wood}}]{kjr+11}
{Keith}, M.~J., {et~al.} 2011, \mnras, 414, 1292

\bibitem[{{Kerr}(2011)}]{ker11}
{Kerr}, M. 2011, PhD thesis, University of Washington, arXiv/1101.6072

\bibitem[{Manchester {et~al.}(2001)Manchester, Lyne, Camilo, Bell, Kaspi,
  D'Amico, McKay, Crawford, Stairs, Possenti, Morris, \& Sheppard}]{mlc+01}
Manchester, R.~N., {et~al.} 2001, MNRAS, 328, 17

\bibitem[{{Marelli} {et~al.}(2011){Marelli}, {De Luca}, \& {Caraveo}}]{mdc11}
{Marelli}, M., {De Luca}, A., \& {Caraveo}, P.~A. 2011, \apj, 733, 82

\bibitem[{{Muslimov} \& {Harding}(2003)}]{mh03}
{Muslimov}, A.~G., \& {Harding}, A.~K. 2003, \apj, 588, 430

\bibitem[{{Muslimov} \& {Harding}(2004)}]{mh04a}
---. 2004, ApJ, 606, 1143

\bibitem[{{Pletsch} {et~al.}(2011){Pletsch}, {Guillemot}, {Allen}, {Kramer},
  {Aulbert}, {Fehrmann}, {Ray}, {Barr}, {Belfiore}, {Camilo}, {Caraveo},
  {Celik}, {Champion}, {Dormody}, {Eatough}, {Ferrara}, {Freire}, {Hessels},
  {Keith}, {Kerr}, {de Luca}, {Lyne}, {Marelli}, {McLaughlin}, {Parent},
  {Ransom}, {Razzano}, {Reich}, {Saz Parkinson}, {Stappers}, \&
  {Wolff}}]{pga+11}
{Pletsch}, H.~J., {et~al.} 2011, \apj, in press (arXiv/1111.0523)

\bibitem[{Ransom(2001)}]{ran01}
Ransom, S.~M. 2001, PhD thesis, Harvard University

\bibitem[{{Ransom} {et~al.}(2011){Ransom}, {Ray}, {Camilo}, {Roberts}, {{\c
  C}elik}, {Wolff}, {Cheung}, {Kerr}, {Pennucci}, {DeCesar}, {Cognard}, {Lyne},
  {Stappers}, {Freire}, {Grove}, {Abdo}, {Desvignes}, {Donato}, {Ferrara},
  {Gehrels}, {Guillemot}, {Gwon}, {Harding}, {Johnston}, {Keith}, {Kramer},
  {Michelson}, {Parent}, {Saz Parkinson}, {Romani}, {Smith}, {Theureau},
  {Thompson}, {Weltevrede}, {Wood}, \& {Ziegler}}]{rrc+11}
{Ransom}, S.~M., {et~al.} 2011, \apjl, 727, L16

\bibitem[{{Ray} \& {Parkinson}(2011)}]{rp11}
{Ray}, P.~S., \& {Parkinson}, P.~M.~S. 2011, in High-Energy Emission from
  Pulsars and their Systems, ed. {D.~F.~Torres \& N.~Rea}, 37 (arXiv/1007.2183)

\bibitem[{{Ray} {et~al.}(2011){Ray}, {Kerr}, {Parent}, {Abdo}, {Guillemot},
  {Ransom}, {Rea}, {Wolff}, {Makeev}, {Roberts}, {Camilo}, {Dormody}, {Freire},
  {Grove}, {Gwon}, {Harding}, {Johnston}, {Keith}, {Kramer}, {Michelson},
  {Romani}, {Saz Parkinson}, {Thompson}, {Weltevrede}, {Wood}, \&
  {Ziegler}}]{rkp+11}
{Ray}, P.~S., {et~al.} 2011, \apjs, 194, 17

\bibitem[{{Romani} {et~al.}(2011){Romani}, {Kerr}, {Craig}, {Johnston},
  {Cognard}, \& {Smith}}]{rkc+11}
{Romani}, R.~W., {Kerr}, M., {Craig}, H.~A., {Johnston}, S., {Cognard}, I., \&
  {Smith}, D.~A. 2011, \apj, 738, 114

\bibitem[{{Romani} \& {Watters}(2010)}]{rw10}
{Romani}, R.~W., \& {Watters}, K.~P. 2010, \apj, 714, 810

\bibitem[{{Saz Parkinson}(2011)}]{par11}
{Saz Parkinson}, P.~M. 2011, in Radio Pulsars: an Astrophysical Key to Unlock
  the Secrets of the Universe, ed. M.~Burgay, N.~D'Amico, P.~Esposito,
  A.~Pellizzoni, \& A.~Possenti, Vol. 1357 (AIP), 48 (arXiv/1101.3096)

\bibitem[{{Saz Parkinson} {et~al.}(2010){Saz Parkinson}, {Dormody}, {Ziegler},
  {Ray}, {Abdo}, {Ballet}, {Baring}, {Belfiore}, {Burnett}, {Caliandro},
  {Camilo}, {Caraveo}, {de Luca}, {Ferrara}, {Freire}, {Grove}, {Gwon},
  {Harding}, {Johnson}, {Johnson}, {Johnston}, {Keith}, {Kerr},
  {Kn{\"o}dlseder}, {Makeev}, {Marelli}, {Michelson}, {Parent}, {Ransom},
  {Reimer}, {Romani}, {Smith}, {Thompson}, {Watters}, {Weltevrede}, {Wolff}, \&
  {Wood}}]{sdz+10}
{Saz Parkinson}, P.~M., {et~al.} 2010, \apj, 725, 571

\bibitem[{{Verde} {et~al.}(2003){Verde}, {Peiris}, {Spergel}, {Nolta},
  {Bennett}, {Halpern}, {Hinshaw}, {Jarosik}, {Kogut}, {Limon}, {Meyer},
  {Page}, {Tucker}, {Wollack}, \& {Wright}}]{vps+03}
{Verde}, L., {et~al.} 2003, \apjs, 148, 195

\bibitem[{{Weltevrede} \& {Johnston}(2008)}]{wj08}
{Weltevrede}, P., \& {Johnston}, S. 2008, \mnras, 391, 1210

\end{thebibliography}
\end{document}